\magnification=1200
\vsize=187mm
\hsize=125mm
\hoffset=4mm
\voffset=10mm
%XXXXXXXXXXXXXXXXXXXXXXXXXXXXXXXXXXXXXXXXXXXXXXXXXXXXXXXXXXXXXXXXXXXXXXXXXXXXXXXXXXXX

\abovedisplayskip=4.5pt plus 1pt minus 3pt
\abovedisplayshortskip=0pt plus 1pt
\belowdisplayskip=4.5pt plus 1pt minus 3pt
\belowdisplayshortskip=2.5pt plus 1pt minus 1.5pt
\smallskipamount=2pt plus 1pt minus 1pt
\medskipamount=4pt plus 2pt minus 1pt
\bigskipamount=9pt plus 3pt minus 3pt

%XXXXXXXXXXXXXXXXXXXXXXXXXXXXXXXXXXXXXXXXXXXXXXXXXXXXXXXXXXXXXXXXXXXXXXXXXXXXXXXXXXXXXXXX

%XXXXXXXXXXXXXXXXXXXXXXXXXXXXXXXXXXXXXXXXXXXXXXXXXXXXXXXXXXXXXXXXXXXXXXXXXXXXXXXXXXXXXXXX
%                   Les Hauts de Pages
%XXXXXXXXXXXXXXXXXXXXXXXXXXXXXXXXXXXXXXXXXXXXXXXXXXXXXXXXXXXXXXXXXXXXXXXXXXXXXXXXXXXXXXXX

\newif\ifpagetitre          \pagetitretrue
\newtoks\hautpagetitre     \hautpagetitre={\hfil}
\newtoks\baspagetitre     \baspagetitre={\hfil\tenrm\folio\hfil}

\newtoks\auteurcourant     \auteurcourant={\hfil}
\newtoks\titrecourant     \titrecourant={\hfil}
\newtoks\chapcourant     \chapcourant={\hfil}

\newtoks\hautpagegauche     \newtoks\hautpagedroite
\hautpagegauche={\vbox{\it\noindent\the\chapcourant\hfill\the\auteurcourant\hfill
{ }\smallskip\smallskip\vskip 2mm\line{}}}
\hautpagedroite={\vbox{\hfill\it\the\titrecourant\hfill{ }
\smallskip\smallskip\vskip 2mm\line{}}}

\newtoks\baspagegauche     \newtoks\baspagedroite
\baspagegauche={\hfil\tenrm\folio\hfil}
\baspagedroite={\hfil\tenrm\folio\hfil}

\headline={\ifpagetitre\the\hautpagetitre
\else\ifodd\pageno\the\hautpagedroite
\else\the\hautpagegauche\fi\fi}

\footline={\ifpagetitre\the\baspagetitre
\global\pagetitrefalse
\else\ifodd\pageno\the\baspagedroite
\else\the\baspagegauche\fi\fi}

%XXXXXXXXXXXXXXXXXXXXXXXXXXXXXXXXXXXXXXXXXXXXXXXXXXXXXXXXXXXXXXXXXXXXXXXXXXXXXXXXXXXXXXX

\auteurcourant={}
\titrecourant={}

%XXXXXXXXXXXXXXXXXXXXXXXXXXXXXXXXXXXXXXXXXXXXXXXXXXXXXXXXXXXXXXXXXXXXXXXXXXXXXXXXXXXXXX
%XXXXXXXXXXXXXXXXXXXXXXXXXXXXXXXXXXXXXXXXXXXXXXXXXXXXXXXXXXXXXXXXXXXXXXXXXXXXXXXXXXXXXX
%                                 MACROS PERSONNELLES
%XXXXXXXXXXXXXXXXXXXXXXXXXXXXXXXXXXXXXXXXXXXXXXXXXXXXXXXXXXXXXXXXXXXXXXXXXXXXXXXXXXXXXX
%XXXXXXXXXXXXXXXXXXXXXXXXXXXXXXXXXXXXXXXXXXXXXXXXXXXXXXXXXXXXXXXXXXXXXXXXXXXXXXXXXXXXX

\font\timeonze=cmr10 scaled 1100

\font\bfonze=cmbx10 scaled 1100

\def\NN{{\mathord{I\!\! N}}}
\def\RR{{\mathord{I\!\! R}}}

\def\CC{{\mathord{C\mkern-16mu{\phantom t\vrule}{\phantom o}}}}

\def\bar#1{{\overline{#1}}}
\def\Rp{\RR^+}
\def\pro#1{{(#1_t)}_{t\geq 0}}
\def\norme#1{\left\vert\left\vert #1\right\vert\right\vert}
\def\normca#1{{\left\vert\left\vert #1\right\vert\right\vert}^2}
\def\ab#1{\left\vert #1\right\vert}

\def\titredeux#1#2#3{\centerline{\bfonze#1}\medskip
     \centerline{\bfonze#2}\bigskip\bigskip\bigskip\centerline{\timeonze#3}}

\def\spa#1#2{\bigskip\medskip\noindent{\bfonze #1\ #2}\par\nobreak\bigskip}
\def\sspa#1#2{\bigskip\noindent{\bf #1\ #2}\par\nobreak\smallskip}

\def\th#1{\bigskip\smallskip\noindent{\bf Theorem #1}$\,$--$\,$}

\def\thl#1#2{\bigskip\smallskip\noindent{\bf Theorem #1} #2$\,$--$\,$}
\def\prp#1{\bigskip\smallskip\noindent{\bf Proposition #1}$\,$--$\,$}

\def\co#1{\bigskip\smallskip\noindent{\bf Corollary #1}$\,$--$\,$}

\def\prf{\bigskip\noindent{\bf Proof}\par\nobreak\smallskip}

\def\ecarte{\vphantom{\buildrel\bigtriangleup\over =}}
\def\findem{\hfill\hbox{\vrule height 2.5mm depth 0 mm width 2.5 mm}}
\def\indic{{\mathop{\rm 1\mkern-4mu l}}}

\def\di{\displaylines}
\def\eq{\eqalignno}
\def\hf{\hfill}
\def\wh{\widehat}
\def\wt{\widetilde}

\def\ld{\ldots}
\def\cd{\cdot}

\def\qed{\findem}

\def\qq{\qquad}

\def\frac#1#2{{{#1}\over{#2}}}
\def\seq#1{{(#1_n)}_{n\in\NN}}

\def\rB{{\cal B}}\def\rD{{\cal D}}
\def\rE{{\cal E}}\def\rF{{\cal F}}\def\rH{{\cal H}}
\def\rL{{\cal L}}
\def\rP{{\cal P}}
\def\rS{{\cal S}}

\def\b{\beta}
\def\g{\gamma}
\def\d{\delta}
\def\e{\varepsilon}

\def\l{\lambda}
\def\s{\sigma}

\def\r{\rho}

\def\x{\chi}

\def\o{\omega}

\def\G{\Gamma}

\def\O{\Omega}

\def\F{\Phi}

\def\TF{T \! \Phi}

%operateurs et integrandes

\def\tr{\hbox{tr}}

\def\NNE{\NN^*}
\def\TF{T \! \Phi}      

\titredeux{THE LANGEVIN EQUATION}{FOR A QUANTUM HEAT BATH}{St\'ephane
  ATTAL${ }^{{ }^1}$ \& Alain JOYE${}^{{\,}^2}$}
\bigskip\bigskip
\centerline{\sevenrm ${}^1$ Institut C. Jordan}
\vskip -.1cm
\centerline{\sevenrm Universit\'e C. Bernard, Lyon 1}
\vskip -.1cm
\centerline{\sevenrm 21, av Claude Bernard}
\vskip -.1cm
\centerline{\sevenrm 69622  Villeurbanne Cedex}
 \vskip -.1cm
\centerline{\sevenrm France}
\smallskip
\centerline{\sevenrm ${}^2$ Institut  Fourier}
\vskip -.1cm
\centerline{\sevenrm Universit\'e de Grenoble 1}
\vskip -.1cm 
\centerline{\sevenrm 100, rue des Maths, BP 74}
\vskip -.1cm
\centerline{\sevenrm  38402 St Martin d'Heres}
\vskip -.1cm
\centerline{\sevenrm France}

\spa{}{Abstract}
{\sevenrm We compute the quantum Langevin equation (or quantum
stochastic differential equation) representing the action of a  quantum
heat bath at thermal equilibrium on a simple quantum system. These
equations are obtained by taking the continuous limit of the 
Hamiltonian description for repeated quantum interactions with a sequence of
photons at a given density matrix state. In particular we specialise
these equations to the case of  thermal equilibrium states. In the
process, new quantum noises are appearing: thermal quantum
noises. We discuss the mathematical properties of these thermal quantum
noises. We compute the Lindblad generator  
associated with the action of the heat bath on the small system. We
exhibit the
typical Lindblad generator that provides thermalization of a given
quantum system.}

\spa{I.}{Introduction}

The aim of Quantum Open System theory (in mathematics as well as in
physics) is to study the interaction of simple quantum systems
interacting with very large ones (with infinite degrees of freedom). In
general the properties that one is seeking  are to exhibit the
dissipation of the small system in favor of the large one, to identify
when this interaction  gives rise to a return to equilibrium or a
thermalization of the small system.

There are in general two ways of studying those system, which usually
represent distinct groups of researchers (in mathematics as well as in
physics). 

The first approach is Hamiltonian. The complete quantum
system formed by the small system and the reservoir is studied
through a Hamiltonian describing the  free evolution of each component and
the interaction part. The associated unitary group gives rise to a
group of *-endomorphisms of a certain von Neumann algebra of
observables. Together with a state for the whole system, this
constitutes a quantum dynamical system. The aim is then to
study the ergodic properties of that quantum dynamical system. This
can be performed via the spectral study of a particular generator of
the dynamical system: the standard Liouvillian. This is the only
generator of the 
quantum dynamical system which  stabilizes the
self-dual cone of the associated Tomita-Takesaki modular
theory. It has the property to encode in its spectrum the ergodic
behavior of the quantum dynamical system.
 Very satisfactory recent results in
that direction were obtained by 
Jaksic and Pillet ([JP1], [JP2] and [JP3]) who rigorously proved the return to
equilibrium for Pauli-Fierz systems, using these techniques.

The second approach is Markovian. In this approach one gives up the
idea of modelizing 
the reservoir and concentrates on the effective dynamics of the small
system. This evolution is supposed to be described by a semigroup of
completely positive maps. These semigroups are well-known and, under
some conditions, admit a generator which is of {\it Lindblad form}:
$$
\rL(X)=i[H,X]+{1\over 2}\sum_i (2L_i^*XL_i-L_i^*L_iX-X L_i^*L_i).
$$
The first order part of $\rL$ represents the usual quantum dynamic
part, while the second 
order part of $\rL$ carries the dissipation. This form has to be compared with
the general form, in classical Markov process theory, of a Feller  diffusion
generator: a first order differential part which carries the classical
dynamics and a second order differential part which represents the
diffusion.
For classical diffusion, such a semigroup
can be realized as resulting of a stochastic differential
equation. That is, a perturbation of an ordinary differential equation
by classical noise terms such as a Brownian motion usually. In
our quantum context,
one can add to the small  system an adequate Fock space which
carries {\it quantum noises} and show that the effective dynamics we have
started with is resulting of a unitary evolution on the coupled system,
driven by a quantum Langevin equation. That is, a perturbation of a
Schr\"odinger-type equation by quantum noise terms.
\bigskip
Whatever the approach is, the study of the action of quantum thermal
baths is  of major importance and has many applications. In the
Hamiltonian approach, the model for such a bath is very well-known
since Araki-Woods' work ([A-W]). But in the Markovian context, it was
not so clear what the correct quantum Langevin equation should be to
 account for the action of a thermal bath. Some equations have been
proposed, in particular by Lindsay and Maassen ([L-M]). But no true physical
justification of them has ever been given. Besides, it is not so clear
what a ``correct'' equation should mean? 

A recent work of Attal and Pautrat ([AP1]) is a good candidate to
 answer that problem. Indeed, consider the setup of a
quantum system (such as an atom) having repeated interactions, for a
short duration 
$\tau$, with elements of a sequence of identical quantum systems (such as
a sequence of photons). The Hamiltonian evolution of such a dynamics
can be easily described. It is shown in [AP1] that in the continuous
limit ($\tau\rightarrow 0$), this Hamiltonian evolution spontaneously
converges to a quantum Langevin equation. The coefficient of the
equation being easily computable in terms of the original
Hamiltonian. 
This work has two interesting consequences: 

-- It justifies the Langevin-type
equations for they are obtained without any probabilistic assumption,
directly from a Hamiltonian  evolution; 

--
It is an effective theorem in the sense that, starting with a 
naive model for a quantum field (a sequence of photons interacting one
after the other with the small system), one obtains explicit quantum Langevin equations which
meet all the usual models of the litterature.

It seems thus natural  to apply this approach in order to derive
the correct quantum Langevin equations for a quantum heat bath. This
is the aim of this article.
\bigskip
We consider a simple quantum system in interaction with a toy model
for a heat bath. The toy model consists in a chain of independent
photons, each of which  in the thermal Gibbs state at inverse
temperature $\b$, which are interacting one after the other with the
small system. Passing to the continuous interaction limit, one
should obtain the correct Langevin equation.

One difficulty here is that in [AP1],
the state of each photon needed to be a pure state (this choice is
crucial in their construction). This is clearly not the case for a
Gibbs state. We solve this problem by taking the 
G.N.S. (or cyclic) representation associated to that state. If the
state space of one 
(simplified) photon was taken to be $n$-dimensional, then taking the
G.N.S. representation  brings us  into a
$n^2$-dimensional space. This may seem far too big and give the
impression we will need too many quantum noises in our model. But we
show  that, in all cases, only $2n$ chanels of noise resist to the passage
to the limit and that they can be naturally coupled two by two to give rise to
$n$ ``thermal quantum noises''. The Langevin equation then remains driven by
$n$ noises (which was to be expected!) and the noises are shown to be
actually  Araki-Woods representations of the usual quantum
noises. Furthermore, the Langevin equation we obtain is very similar
 to
the model given in [L-M].

Altogether this confirms we have identified the correct Langevin
equation modelizing the action of a quantum heat bath.

An important point to notice is that our construction does not
actually use the fact that the state is a Gibbs-like state, it is
valid for any density matrix. 
\bigskip
This article is organized as follows. In section II we present the toy
model for the bath and the Hamiltonian description of the repeated
interaction procedure. In section III we present the Fock space, its
quantum noises, its approximation by the toy model and 
the main result of [AP1]. In section IV we detail the
G.N.S. representation of the  bath and compute the unitary operator,
associated with the 
total Hamiltonian, in that representation. In section V, applying the
continuous limit procedure we derive the limit quantum langevin
equation. In the process, we identify particular quantum noises that
are naturally appearing and baptize them  ``thermal quantum
noises'', in the case of a heat bath. The properties of those thermal
quantum noises are studied 
in section VI; in particular we justify their name. In section VII,
tracing out the noise, we compute the Lindblad generator of the induced
semigroup on the small system. In section VII, being given any finite
dimensional quantum
system with its Hamiltonian, we show how to
construct a Lindblad generator, representing some interaction with a
heat bath, such that the quantum system thermalizes.

\spa{II.}{The toy model}
We  describe here the physical model of repeated interactions with the
 bath toy model.
\bigskip
The quantum system (we shall often call ``small system'') to be put in 
contact with the bath is
represented by a separable Hilbert  space $\rH_S$ , as  state space,
and 
a self-adjoint operator $H_S$, as  Hamiltonian.

The toy model for the heat bath is the chain 
$$
\bigotimes_{k\in\NNE} \CC^{N+1}
$$
of copies of $\CC^{N+1}$, where $N\geq 1$ is a fixed integer. Each
copy of $\CC^{N+1}$ represents the (simplified) state space of a
photon. By this countable tensor product we mean the following. We
consider a fixed orthonormal basis $\{e_0,e_1,\ld,e_N\}$ 
of $\CC^{N+1}$, corresponding to the eigenstates of the photon ($e_0$
being the ground state); the countable tensor product is taken with
respect to the ground state $e_0$. Together with this structure we
consider the associated basic matrices
$a^i_j$, $i,j=1,\ld,N$, acting on $\CC^{N+1}$ by 
$$
a^i_j\,e_k=\d_{ik}\,e_j
$$
and their natural ampliations to $\otimes_{k\in\NNE} \CC^{N+1}$ given
by
$$
a^i_j(k)=\cases{a^i_j& on the $k$-th copy of $\CC^{N+1}$\cr
I& on the other copies.\cr}
$$
The Hamiltonian of one photon  is the operator
$$
H_{R}=\sum_{i=0}^N\g_i\, a^0_ia^i_0,
$$
where the $\g_i$'s are real numbers. Here notice two points. 

We have
assumed the Hamiltonian $H_R$ to be diagonal in the chosen basis. This
is of course not actually a true  restriction, for one can 
always choose such a basis.Note that $H_R$ describe the total energy of a single photon, not the
whole field of photon. For this we differ from the model studied in [AJ1]. 

The second point is
that $\g_0$ is the ground state eigenvalue, it should then be smaller
than the other $\g_i$. One usually assumes that it is equal to 0, but
this is not actually necessary in our case, we thus do not specify its
value. The only hypothesis we shall make here is that $\g_0<\g_i$, for
all $i=1,\ld N$. This hypothesis means that the ground eigenspace is
simple, it is not actually a necessary assumption, it only simplifies
our discussion. At the end of  section V we discuss what 
changes if we leave out this hypothesis. 

Finally, notice that the
other eigenvalues $\g_i$ need not be simple in our discussion. 

\bigskip
When the system and a photon are interacting, we consider the
state space $\rH_S\otimes\CC^{N+1}$ together with the interaction
hamiltonian
$$
H_I=\sum_{i=1}^N\left(V_i\otimes a^0_i+V^*_i\otimes a^i_0\right), 
$$
where the $V_i$'s are bounded operators on $\rH_S$. This is a usual 
dipole-type interaction Hamiltonian. The total 
Hamiltonian for the small system and one photon is thus 
$$
H=H_S\otimes I+I\otimes H_{R}+\sum_{i=1}^N \left(V_i\otimes a^0_i+V_i\otimes
a^i_0\right).
$$
Finally, the state of each photon is fixed to be given by a density
matrix $\r$ which is a function of $H_R$. We have in mind the usual
thermal Gibbs state at 
inverse temperature $\b$ :
$$
\r_{\b}={1\over Z}\, e^{-\b H_{R}},
$$
where $Z=\tr(e^{-\b H_{R}})$, but our construction applies to more
general states $\r$. 

Note that $\r_\b$ is also diagonal in our orthonormal basis. Its diagonal
elements are denoted by $\{\b_{0},\b_{1},\ld,\b_{n}\}$. 
\bigskip
We shall now describe the repeated  interactions of the system
$\rH_S$ with the 
chain of photons. The system $\rH_S$ is first in contact
with the first photon only and they interact together according to the above
Hamiltonian $H$. This lasts for a time length $\tau$. The
system $\rH_S$ then stops interacting with the first photon and starts 
interacting with the second photon only. This second interaction is
directed by the same  Hamiltonian $H$ on the corresponding spaces and
it lasts for the same duration $\tau$, and so on... This is mathematically
described as follows. 

On the space $\rH_S\otimes \CC^{N+1}$, consider the unitary
operator representing the coupled evolution during the time interval
$[0,\tau]$:
$$
U=e^{-i\tau H}.
$$
This single interaction is therefore described in the Schr\"odinger
picture by
$$
\r\mapsto U\,\r\,U^\ast
$$
and  in the Heisenberg picture by
$$ X \mapsto U^\ast X U.$$
After this first interaction, we repeat it but coupling
the same $\rH_S$ with a new copy of $\CC^{N+1}$. This means that this new
copy was kept isolated until then; similarly the previously considered
copy of $\CC^{N+1}$ will remain isolated for the rest of the experience. 

The sequence of interactions can be described in the following way:
the state space for the whole system is
$$
\rH_S\otimes\bigotimes_{\NN^*}\CC^{N+1}.
$$
Consider the unitary operator $U _k$ which acts as $U$ on the
tensor product of $\rH_S$ and the $k$-th copy of $\CC^{N+1}$, and
which acts as the identity on all the other copies of $\CC^{N+1}$.

The effect of the $k$-th interaction in the Schr\"odinger picture is
$$ \r \mapsto U _k \, \r\,U _k ^*,$$
for every density matrix $\r$ on $\rH_S\otimes_\NNE \CC^{n+1}$.  In particular
 the effect of the $k$ first 
interactions is 
$$ \r \mapsto V_k\, \r\, V_k^*$$
where $V_k=U_kU_{k-1}\ld U_1$.
\bigskip
Such a Hamiltonian description of the repeated interaction
procedure has no chance to give any non-trivial limit in the
continuous limit ($\tau\rightarrow 0$) without asking a certain
renormalization of the interaction. This renormalization can be thought of
as  making the Hamiltonian 
 depend on $\tau$, or can be also seen as renormalizing the field
operators $a^0_j, a^i_0$ of the photons.
As is shown is [AP1] (see the detailed discussion in section III), for our
repeated interaction model 
to give rise 
to a Langevin equation in the limit, we need the interaction part 
of the Hamiltonian to be affected by a weight $1/\sqrt\tau$. Hence, from
now on, the total Hamiltonians we shall consider on $\rH_S\otimes\CC^{N+1}$ are
$$
H=H_S\otimes I+I\otimes H_{R}+{1\over{\sqrt\tau}}\sum_{i=1}^N
\left(V_i\otimes a^0_i+V_i^*\otimes 
a^i_0\right).\eqno{(1)}
$$
In [AJ1], one can find a discussion about this time renormalization
and its interpretation in terms of weak coupling limit for repeated
quantum interactions.

\spa{III.}{The continuous limit setup}

We present here all the elements of the continuous limit result:
the structure of the corresponding Fock space, the quantum noises, the
approximation of the Fock space by the photon chain and [AP1]'s main theorem.

\sspa{III.1}{The continuous tensor product structure}

First, as a guide to intuition, let us make more explicit the
structure of the photon chain. We let $\TF$ denote the tensor product
$\otimes _{\NN^*} \CC^{N+1}$ with respect 
to the stabilizing sequence $e_0$. This simply means
 that an orthonormal basis of $\TF$ is given by the family
 
$$ \{ e_\s; \s \in \rP_{\NNE,N}\}$$
where

-- the set $\rP _{\NN, N}$ is  the set of finite subsets 
$$ \{ (n_1, i _1), \ldots, (n_k, i_k)\}$$
of $\NNE \times \{1,\ldots , N\}$ such that the $n_i$'s are mutually different;

-- $e_\s$ denotes the vector 
$$
\O\otimes\ld\otimes\O\otimes e_{i_1}\otimes
\O\otimes\ld\otimes\O\otimes e_{i_2}\otimes\ld
$$
where $e_{i_1}$ appears in $n_1$-th copy of $\rH$, where $e_{i_2}$
appears in $n_2$-th copy of $\rH$... Here $\O$ plays the same role as
$e_0$ in the toy model.

This is for a vector basis on $\TF$. From the point of view of
operators, we denote by $a^i_j(k)$ the natural ampliation of the
operator $a^i_j$ 
to $\TF$ which acts on the copy number $k$ as $a^i_j$ and the identity
elsewhere. That is, in terms of the basis $e_\s$, 
$$
a^i_j(k)e_\s=\indic_{(k,i)\in \s}\,e_{(\s\setminus (k,i))\cup (k,j)}
$$
if neither $i$ nor $j$ is zero, and
$$
\eqalign{
a^i_0 (k) e_\s &= \indic_{(k,i) \in \s}\, e_{\s \setminus (k,i)}, \cr
a^0_j(k)e_\s &=\indic_{(k,0) \in \s}\, e_{\s\cup (k,j)},\cr
a^0_0 (k) e_\s &= \indic_{(k,0) \in \s}\, e_{\s},\cr}
$$
where $(k,0)\in \s$ actually means ``for any $i$ in $\{1,\ld,N\}$,
$(k,i)\not\in\s$''.
\bigskip
We now describe the structure of the continuous version of
the chain of photons. The  structure we are going to present here
is rather original and not  much expanded in the literature. It
is very different from the usual presentation of quantum stochastic
calculus ([H-P]), but it actually constitutes a very natural
language for our purpose: approximation of the atom field by atom
chains. This approach is taken from [At1]. We first start with a
heuristic discussion. 

By a continuous version of the atom chain $T\F$ we mean  a Hilbert
space with a structure which makes it 
the space  
$$
\F=\bigotimes_{\RR^+}\CC^{N+1}.
$$
We have to give a meaning to the above notation. This could be
achieved by invoquing  the framework of continous tensor products of Hilbert
spaces (see [Gui]), but we prefer to give a self-contained
presentation which fits better with our approximation procedure.

Let us make out an
idea of what it should look like by mimicking, in a 
continuous time version, what we have described in $T\F$. 

The countable orthonormal basis $e_\s, \s\in\rP_{\NNE, N}$ is replaced by
a continuous orthonormal  basis $d\x_\s,\, \s\in\rP_{\Rp, N}$, where
$\rP_{\Rp, N}$ is the 
set of finite subsets of $\RR^+\times \{1,\ld,N\}$. With the same idea
as for $\TF$, 
this means that each copy of $\CC^{N+1}$ is equipped with an orthonormal
basis $\{\O,d\x^1_t, \ld,d\x^N_t\}$ (where $t$ is the parameter
attached to the copy we 
are looking at).

Recall the representation of an element $f$ of $T\F$:
$$
\eqalign{
f&=\sum_{\s\in\rP_{\NNE, N}}  f(\s)\, e_\s,\cr
\normca f&=\sum_{\s\in\rP_{\NNE,N}} \! \ab{f(\s)}^2,\cr
}
$$
it is  replaced by an integral version of it in $\F$:
$$
\eq{
f&=\int_{\rP_{\Rp,N}} \! f(\s)\, d\x_\s,\cr
\normca f&=\int_{\rP_{\Rp,N}} \! \ab f^2\, d\s.\cr
}
$$
This last integral needs to be explained: the measure $d\s$ is  a
``Lebesgue measure'' on $\rP_{\Rp,N}$, as  will be explained later.

From now on, the notation $\rP$ will denote, depending on the context,
the set $\rP_{\NNE,N}$ or $\rP_{\Rp,N}$. 

A good basis of operators acting on $\F$ can be obtained by mimicking
the operators $a^i_j(k)$ of $\TF$. We have here a set of
infinitesimal operators $da^i_j(t)$, $i,j \in\{0,1,\ld,N\}$,  acting
on the ``t-th" copy of 
$\CC^{N+1}$ by:
$$
\eq{
da^0_0(t)\, d\x_\s&=d\x_{\s}\, dt\, \indic_{t\not\in \s}\cr
da^0_i(t)\, d\x_\s&=d\x_{\s\cup \{(t,i)\}}\, \indic_{t\not\in \s}\cr
da^i_0(t)\, d\x_\s&=d\x_{\s\setminus \{(t,i)\}}\, dt\, \indic_{(t,i)\in \s}\cr
da^i_j(t)\, d\x_\s&=d\x_{(\s\setminus\{(t,i)\})\cup\{(t,j)\}}\,
\indic_{(t,i)\in \s}\cr 
}
$$
for all $i,j \in \{1,\ld,N\}$. We shall now describe a rigourous setup
for the above heuristic discussion. 
\bigskip
\def\pcc{\rP}
\def\rb{\RR}
\def\cb{\CC}
\def\pck{\rP_k}
\def\fc{\rF}
\def\vi{\emptyset}
We recall the structure of the bosonic Fock space $\F$ and its basic
structure (cf [At1] for more details and [At2] for a complete study of
the theory and its connections with classical stochastic processes).

Let $\F=\G_s(L^2(\Rp, \CC^N))$ be the symmetric (or bosonic) Fock space
over the space $L^2(\Rp, \CC^N)$. 
We shall  give here a very efficient presentation of that space, the
so-called {\it Guichardet interpretation} of the Fock space. 

Let $\pcc$ ($=\pcc_{\Rp, N}$) be the set of finite subsets
$\{(s_1,i_1),\ld,(s_n,i_n)\}$ of
$\rb^+\times \{1,\ld,N\}$ such that the $s_i$ are two by two
different. Then $\pcc = \cup_k \pck$ where $\pck$ is the subset of
$\pcc$ made of $k$-elements subsets of 
$\ \rb^+\times \{1,\ld,N\}$. By ordering the $\Rp$-part of the
elements of 
$\s\in\pck$, the set $\pck$ can be identified with the increasing simplex
$\Sigma _k = \{0<t_1 < \cdots <t_k\}\times \{1,\ld,N\}$ of
$\rb^k\times \{1,\ld,N\}$. Thus $\pck$ 
inherits a measured space structure from the Lebesgue measure on
$\ \rb^k$ times the counting measure on $\{1,\ld,N\}$. This also gives
a measure 
structure on $\pcc$ if we specify 
that on $\pcc_0 = \{\vi\}$ we put the measure $\delta
_\vi$. Elements of $\pcc$ are often denoted by $\sigma $, the measure on
$\pcc$ is denoted by $d\sigma $. The $\sigma $-field obtained this way on
$\pcc$ is denoted by $\fc$.

We identify any element $\s\in\rP$ with a family
$\{\s_i, \,  i\in \{1,\ld,N\} \}$ of (two by two disjoint)  subsets of
$\Rp$ where  
$$
\s_i=\{s\in\Rp; (s,i)\in\s\}.
$$

\def\Unn{\indic}

The {\it Fock space\/} $\Phi$ is the space $L^2(\pcc,\fc,d\sigma
)$. An element $f$ of $\Phi$ is thus a measurable function $f:\pcc \to
\cb$ such that
$$
\normca{f} = \int_{\pcc} |f(\sigma )|^2\ d\sigma  < \infty.
$$
Finally, we put $\O$ to be the {\it vacuum vector} of $\Phi$, that is,
$\O(\s)=\d_\emptyset(\s)$.
\bigskip
One can define, in the same way, $\pcc_{[a,b]}$ and $\Phi_{[a,b]}$ by
replacing $\rb^+$ with $[a,b]\subset \rb^+$. There is a natural
isomorphism between $\Phi_{[0,t]} \otimes  \Phi_{[t,+\infty[}$ and
$\Phi$ given by 
$h\otimes g \mapsto f$ where $f(\sigma ) = h(\sigma \cap [0,t]) g
(\sigma  \cap [t,+\infty[)$. This is, with our notations, the usual
exponential property of Fock spaces. Note that in the sequel we
identify $\Phi_{[a,b]}$ with a subspace of $\Phi$, the subspace 
$$
\{f\in\Phi; f(\s)=0\ \hbox{unless }\s\subset [a,b]\}.
$$

We now define a particular family of curves in $\F$, which is going to be of
great importance here. Define $\chi^i _t \,{\in}\, \Phi$ by
$$
\chi^i _t(\sigma ) = \cases {
\Unn_{[0,t]}(s) &if~~$\sigma = \{(s,i)\} $\cr 0 &otherwise. }
$$
Then  notice that for all $t\in\Rp$ we have that $\chi^i _t$ belongs
to $\Phi_{[0,t]}$. We actually have much more than that: 
$$\chi^i_t -
\chi^i_s \, \in \,  \Phi_{[s,t]} \hbox{\ for all\ } s\leq t.
$$
This last property can be checked immediately from the definitions, and
it is going to be of great importance in our construction. Also notice
that $\x^i_t$ and $\x^j_s$ are orthogonal elements of $\F$ as soon as $i\not
=j$. One can show that, apart from trivialities, the curves
$\pro {\x^i}$ are the only ones to share these properties. 
\bigskip
These properties allow to define the so-called {\it Ito integral\/} on
$\Phi $. Indeed,  let $g=\{(g^i_t)_{t\ge
0}, \, i \in\{1,\ld,N\}\}$ be families of elements of $\ \Phi $
indexed by both $\RR_+$ and $\{1,\ld,N\}$, such that 
\smallskip
 i) $t\mapsto \|g^i_t\|$ is measurable, for all $i$,
\smallskip
ii) $g^i_t \,{\in}\, \Phi _{[0,t]}$ for all $t$,
\smallskip
iii) $\sum_{i=1}^N\int^\infty _0 \|g^i_t\|^2\ dt < \infty $,
\smallskip
\def\nb{\NN}
\noindent then one says that $g$ is {\it Ito integrable}
and we define its {\it Ito integral} 
$$
\sum_{i=1}^N\int^\infty _0 g^i_t\ d\chi^i_t
$$
to be the limit in $\Phi$ of
$$
\sum_{i=1}^N \sum_{j\in\NN} \wt g^i_{t_j}
\otimes \left(\chi^i _{t_{j+1}} - \chi^i_{t_j}\right)\eqno(2)
$$
where $\rS=\{t_j,~j{\in} \nb\}$ is a partition of $\rb^+$ which is
understood to 
be refining and to have its diameter tending to $0$, and $(\wt g^i_\cd)_i$
is an Ito integrable family in $\F$, such that for each $i$, $t\mapsto
\wt g^i_t$ is a step process, and which converges to $(g^i_\cd)_i$ in 
$L^2(\Rp\times\rP)$.

Note that by assumption we always  have that $\wt g^i_{t_j}$ belongs to
$\F_{[0,t_j]}$ and $\x^i_{t_{j+1}}-\x^i_{t_j}$ belongs to $\F _{[t_j,
t_{j+1}]}$, hence the tensor product symbol in (2).  

Also note that, as an example, one can take 
$$
\wt g^i_t = \sum_{t_j\in\rS}{1\over t_{j+1}-t_j} \int^{t_{j+1}}_{t_j}
P_{t_j} g^i_s\,ds\, \indic_{[t_j,t_{j+1}[}(t)
$$
if $t\in[t_j,t_{j+1}]$,
where
$P_t$ denotes the orthogonal projection onto $\Phi _{[0,t]}$. 
\bigskip
One then obtains the following properties ({[At1], Proposition 1.4}),
where $\vee \s$ means $\max\{s\in\Rp; (s,k)\in\s$ for some $k\}$ and
where $\s-$ denotes the set $\s\setminus(\vee\s,i)$ if $(\vee\s,i)\in\s$.

\th{1.}{\it The Ito integral $I(g)=\sum_i\int^\infty _0 g^i_t \
d\chi^i_t$, of an Ito integrable family $g=(g^i_\cd)_{i=1}^N$,  is the  element of $\Phi $
given by
$$
I(g)(\s)=\cases{0& if $\s=\emptyset$\cr
g^i_{\vee\s}(\s-)& otherwise.\cr}
$$
It satisfies the {\rm Ito isometry formula}:
$$
\normca{I(g)}=\Big\|{\sum_{i=1}^N\int^\infty _0 g_t^i \ d\chi^i _t}
\Big\| ^2 = \sum_{i=1}^N\int^\infty _0 
\norme{g^i_t}^2\,dt~.\eqno(3) 
$$}
\qed
\bigskip
In particular, consider a family $f=(f_i)_{i=1}^N$ which belongs to $
L^2(\rP_1)=L^2(\Rp\times\{1,\ld,N\})$, 
then the family  $(f_i(t) \O)$, $t\in\Rp$, $i=1,\ld,N$, is clearly
Ito integrable. Computing its Ito integral we find that
$$
I(f)=\sum_{i=1}^N\int_0^\infty f_i(t)\O\, d\x^i_t
$$
is the element of the first particle space of the Fock space $\F$
associated with  the
function $f$, that is,  
$$
I(f)(\s)=\cases{f_i(s)&if $\s=\{(s,i)\}$\cr
0&otherwise.\cr}
$$
\bigskip
Let $f {\in} L^2(\rP_n)$, one can easily define the {\it iterated Ito
integral\/} on $\ \Phi $:
$$
I_n(f) = \int_{\rP_n}f(\s)\, d\chi_\s
$$
by iterating the definition of the Ito integral:
$$
I_n(f)=\sum_{i_1,\ld,i_n \in \{1,\ld,N\}}\int_0^\infty\int_0^{t_n}\ld\int_0^{t_2}
f_{i_1,\ld, i_n}(t_1,\ld,t_n)\O\,\,d\x^{i_1}_{t_1}\, \ld\,
d\x^{i_n}_{t_n}.
$$
We obtain this way an element of $\F$ which is actually the
representant of $f$ in the $n$-particle subspace of $\F$, that is
$$
[I_n(f)](\s)=\cases{f_{i_1,\ld,i_n}(t_1,\ld,t_n)&if
$\s=\{(t_1,i_1)\cup\ld\cup(t_n,i_n)\}$\cr0&otherwise.\cr}
$$
Finally, for any $f\in \rP$ we put
$$
\int_\rP f(\s)\, d\x_\s
$$
to denote the series of iterated Ito integrals
$$
f(\emptyset)\O+\sum_{n=1}^\infty\, \sum_{i_1,\ld,i_n
=1}^N\int_0^\infty\int_0^{t_n}\ld\int_0^{t_2} 
f_{i_1,\ld, i_n}(t_1,\ld,t_n)\O\, d\x^{i_1}_{t_1}\, \ld\,
d\x^{i_n}_{t_n}.
$$
We then have the following representation ([At1], Theorem 1.7).

\thl{2.}{[Fock space chaotic representation
property]}{\it Any element 
$f$ of $\Phi $ admits a {\rm 
Fock space chaotic representation}
$$f = \int_{\pcc} f(\sigma )\ d\chi _\sigma \eqno{(4)}$$
satisfying the isometry formula
$$\|f\|^2 = \int_{\pcc} |f(\sigma )|^2\ d\sigma. \eqno{(5)}$$
This representation is unique.}
\bigskip
The above theorem is the exact expression of the heuristics we wanted
in order to describe the space 
$$
\F=\bigotimes_{\Rp}\rH.
$$
Indeed, we have, for each $t\in\Rp$,  a family of elementary
orthonormal elements $\{\O,d\x^1_t,\ld,d\x^N_t \}$
(a basis of $\rH$) whose (tensor) products
$d\x_\s$ form a continuous 
basis of $\F$ (formula (4)) and, even more, form an orthonormal
continuous basis (formula (5)).

\sspa{III.2}{The quantum noises}

The space $\F$ we have constructed is the natural space for defining
{\it quantum noises}. These quantum noises are the natural, continuous-time,
extensions of the basis operators $a^i_j(n)$ we met in the atom chain
$\TF$. 

As indicated in the heuristic discussion above, we shall deal with a
family of infinitesimal operators $da^i_j(t)$ on $\F$ which act on the
continuous basis $d\x_\s$ in the same way as their discrete-time
counterparts $a^i_j(n)$ act on the $e_\s$. The integrated version of the
above 
heuristic infinitesimal formulas easily gives an exact formula for the action
of the operators $a^i_j(t)$ on $\F$:
$$
\eqalignno{
[a^0_i(t)f](\sigma ) &= \sum_{s{\in} \sigma_i \atop s\leq t} f(\sigma
\setminus (s,i)),\cr 
[a^i_0(t)f](\sigma ) &= \int^t_0 f(\sigma \cup  (s,i))\ ds,\cr
[a^i_j(t)f](\sigma ) &= \sum_{s{\in} \sigma_i \atop s\leq t}\
f\left((\sigma\setminus(s,i))\cup(s,j)\right)\cr
[a^0_0(t)f](\s)&=t\, f(\s)\cr
}
$$
for $i,j\not=0$.
\bigskip
All these operators, except $a^0_0(t)$, are unbounded, but 
note that a good common domain to all of them is
\def\dc{\rD}
$$\dc = \Big\{ f{\in} \Phi ~;~~\int_{\pcc} |\sigma | ~|f(\sigma )|^2\
d\sigma  < \infty  \Big\}~.$$
\bigskip

\def\sup{\hbox{sup}}
This family of operators is characteristic and universal in a  sense
which is close to the one of the curves $\x^i_t$. Indeed, one can
easily check that in the decomposition of
$\F\simeq\F_{[0,s]}\otimes\F_{[s,t]}\otimes\F_{[t,+\infty[}$, the operators
$a^i_j(t)-a^i_j(s)$ are all of the form
$$
I\otimes (a^i_j(t)-a^i_j(s))_{\vert\F_{[s,t]}}\otimes I.
$$
This property is fundamental for the definition of the quantum
stochastic integrals and, in the same way as for $(\x^i_\cd)$,  these operator
families are the only ones to share that property (cf [Coq]).

This property allows to consider Riemann sums:
$$
\sum_k H_{t_k}\left(a^i_j(t_{k+1})-a^i_j(t_k)\right)\eqno{(6)}
$$
where $\rS=\{0=t_0<t_1<\ld<t_k<\ld\}$ is a partition of $\Rp$, where
$\pro H$ is 
 a family of operators on $\F$ such that  
\smallskip
-- each $H_t$ is an operator of the form $H_t\otimes I$ in the tensor
   product space 
$\F=\F_{[0,t]}\otimes\F_{[t,+\infty[}$ (we say that $H_t$ is a {\it $t$-adapted
operator} and that $\pro H$ is an {\it adapted process of operators}), 
\smallskip
-- $\pro H$ is a {\it step process}, that is, it is constant on intervals:
$$
H_t=\sum_kH_{t_k}\indic_{[t_k,t_{k+1}]}(t).
$$
In particular, the operator product
$H_{t_k}\left(a^i_j(t_{k+1})-a^i_j(t_k)\right)$ is actually a tensor
product of operators 
$$
H_{t_k}\otimes\left(a^i_j(t_{k+1})-a^i_j(t_k)\right).
$$
Thus this product is commutative and
does not impose  any new domain constraint on the operators apart from
the ones attached to the operators $H_t$ and
$a^i_j(t_{k+1})-a^i_j(t_k)$ themselves.
The resulting operator associated to the Riemann sum (6) is denoted by
$$
T=\int_0^\infty H_s\, da^i_j(s).
$$
One can compute the action of $T$ on a ``good'' vector $f$ of its
domain and  obtain explicit formulas which are not worth developping
here (cf [At1] for more details).
For general operator processes $\pro H$ (still adapted but not a step process
anymore) and for a general $f$, these explicit formulas can be extended and
they are kept as a
definition for the domain and for the action of the operator 
$$
T=\int_0^\infty H_s\, da^i_j(s).
$$
The maximal domain and the explicit action of the above operator can
be described but  also are not worth developing  here (cf [A-L]). The
main point with these quantum stochastic integrals is that, when
composed, they satisfy a Ito-type integration by part formula. This
formula can be summarized as follows, without taking care at all of
domain constraints. Let 
$$
T=\int_0^\infty H_s\, da^i_j(s),\ 
S=\int_0^\infty K_s\, da^k_l(s).
$$
For every $t\in\Rp$ put 
$$
T_t=\int_0^\infty H_s\indic_{[0,t]}(s)\, da^i_j(s)
$$
and the same for $S_t$. We then have
$$
TS=\int_0^\infty H_sS_s\, da^i_j(s)+\int_0^\infty T_sK_s\, da^k_l(s)+
\int_0^\infty H_sK_s \,\widehat{\d}_{il}\,da^k_j(s),\eqno{(7)}
$$
where 
$$
\widehat{\d}_{il}=\cases{\d_{il}& if $(i,l)\not =(0,0)$\cr
0& if $(i,l)=(0,0)$.}
$$
The last term appearing in this Ito-type formula is often summarized
by saying that the quantum noises satisfy the formal formula:
$$
da^i_j(s) da^k_l(s)=\widehat{\d}_{il}\, da^k_j(s).
$$

\sspa{III.3}{Embedding and approximation by the Toy Fock space}

We now describe the way the chain and its basic operators can be
realized as a subspace of the Fock space and a projection of the
quantum noises. The subspace associated with the atom chain is attached
to the choice of some partition of $\Rp$ in such a way that the
expected properties are satisfied:
\smallskip
-- the associated subspaces increase when the partition refines and they
  constitute an approximation of $\F$ when the diameter of the
  partition goes to 0,
\smallskip
-- the associated basic operators are restrictions of the others when
   the partition increases and they constitute an approximation of the
   quantum noises when the diameter of the partition goes to 0.
\bigskip\def\scc{\rS}
Let $\ \scc = \{0 = t_0 < t_1 < \cdots <t_n < \cdots \}$ be a partition of
$\ \rb^+$ and $\delta (\scc) = \sup_i |t_{i+1}-t_i|$ be the diameter
of $\ \scc$. For $\ \scc$  fixed, define $\Phi _n = \Phi
_{[t_{n-1},t_{n}]}$, $n{\in} \NNE$. We clearly have that  $\Phi $ is
naturally isomorphic to the countable tensor product $ \otimes_{n{\in}
\NNE} \Phi _n$ (which is again understood to be defined with respect to the
stabilizing sequence $(\O)_{n{\in} 
\nb}$).

For all $n{\in} \nb^*$, define for $i,j\in\{1,\ld,N\}$
$$
\eqalign{
e_i(n) &= {\chi^i _{t_{n}}-\chi^i _{t_{n-1}}\over \sqrt
{t_{n}-t_{n-1}}} \in \Phi 
_n~,\cr
a^i_0(n) &= {a^i_0(t_{n})-a^i_0({t_{n-1}})\over
\sqrt{t_{n}-t_{n-1}}}\circ P_{1]},\cr 
a^i_j(n) &=  P_{1]}\circ\left(a^i_j(t_{n})-a^i_j(t_{n-1})\right)\circ
P_{1]},\cr 
a^0_i(n) &= P_{1]} \circ{a^0_i(t_{n})-a^0_i({t_{n-1}})\over
\sqrt{t_{n}-t_{n-1}}}~,\cr
a^0_0(n) &=  P_{0]},\cr
}
$$
where for $i=0,1$ and $P_{i]}$ is the orthogonal projection onto
$L^2(\pcc_{ i})$. The above definitions are understood to be 
valid on $\Phi _n$ only, the corresponding operator acting as
the identity operator $I$ 
on the others $\Phi _m$'s. 
\smallskip
For every $\s\in\rP=\rP_{\NNE,N}$, define $e_\s$ from the $e_i(n)$'s in
 the same way as for $\TF$:
$$
e_\s=\O\otimes\ld\otimes\O\otimes e_{i_1}(n_1)\otimes
\O\otimes\ld\otimes\O\otimes e_{i_2}(n_2)\otimes\ld
$$
in $\otimes_{n\in\NNE}\rH_n$. Define $T\Phi (\scc)$ to
be the space of $f {\in} \Phi $ which are of the form
$$
f =  \sum_{\s{\in} \rP}f(\s) e_\s
$$
(note that the condition $\|f\|^2 = \sum_{\s{\in} \rP} |f(\s)|^2 <
\infty $ is automatically satisfied). The space $T\Phi (\scc)$ can be
clearly  and naturally identified to the spin chain 
$\TF$. 
The space $T\Phi (\scc)$ is a closed subspace of $\ \Phi
$.  We denote by
$P_\rS$ the operator of orthogonal
projection from $\Phi $ onto $T\Phi (\scc)$. 
\smallskip 
The main point is that the above operators $a^i_j(n)$ act on
$T\F(\rS)$ in the same way as the the basic operators of $T\F$ (cf
[AP1], Proposition 8). 

\prp{3.}{\it We have, for all $i,j=1,\ld,N$
$$
\eqalign{
&\cases{a^i_0(n)\,e_j(n)=\d_{ij}\O\cr
a^i_0 \,\O =0}\cr
\noalign{\vskip3pt}
&\cases{a^i_j(n)\, e_k(n)=\d_{ik}e_j(n)\cr
a^i_j \,\O =0}\cr
\noalign{\vskip3pt}
&\cases{a^0_i(n)\, e_j(n)=0\cr
a^0_i(n)\, \O =e_i(n)}\cr
\noalign{\vskip3pt}
&\cases{a^0_0(n)\, e_k(n) = 0 \cr
a^0_0 \,\O = \O.}\cr}
$$
}

Thus the action of the operators $a^i_j$ on the $e_i(n)$ is
exactly the same as the action of the corresponding operators on the spin chain
of section II; the operators $a^i_j(n)$ act on $T\Phi (\scc)$ exactly
in the same way as the 
corresponding operators do on $\TF$. We have completely embedded the toy
Fock space structure into the Fock space.
\bigskip
We are now going to see that the Fock space $\Phi$ and its basic
operators $a^i_j(t)$, $i,j\in\{0,1,\ld, n\}$ can be approached by the toy Fock
spaces $T \Phi(\scc)$ and their basic operators $a^i_j(n)$.
We are given a sequence $(\rS_n)_{n{\in} \NN}$ of partitions which are
getting finer and finer and whose diameter $\delta (\scc_n)$ tends to
$0$ when $n$ tends to $+\infty $. Let  $T
\Phi(n) = T \Phi(\rS_n)$ and $P_n= P_{\rS_n}$,  for all $n{\in}
\NN$. We then have  the following convergence result (see [AP1],
Theorem 10), where the reader needs to recall the domain $\rD$ introduced
in section III.2.

\th{4.}{\it

i) The orthogonal projectors $P_n$ converge strongly  to the identity
operator I on $\F$. That is, any $f\in\F$ can be approached in $\F$ by
a sequence $\seq f$ such that $f_n\in\TF(n)$ for all $n\in\NN$. 
\smallskip
ii) If $\ \scc_n = \{0 = t^n_0 < t^n_1 < \cdots < t^n_k <
\cdots \}$, then for all $t {\in} \rb^+$, all $i,j=1,\ld ,n$ the operators
$$
\eq{
\sum_{k;t^n_k\leq t}& a^i_j(k),\cr
\sum_{k;t^n_k\leq t}&
\sqrt{t^n_{k}-t^n_{k-1}}\, a^i_0(k),\cr
\sum_{k;t^n_k\leq t}&
\sqrt{t^n_{k}-t^n_{k-1}}\, a^0_i(k)\cr
\hbox{and}\ \sum_{k;t^n_k\leq t}&
({t^n_{k}-t^n_{k-1}})\, a^0_0(k)\cr
}
$$ 
converge strongly on $\dc$ to $a^i_j(t)$,
$a^i_0(t)$, $a^0_i(t)$ and $a^0_0(t)$ respectively.
}
\bigskip
We have fulfilled our duties: not only the space $\TF(\rS)$ recreates
$\TF$ and its basic operators as a subspace of $\F$ and a projection
of its quantum noises, but, when $\d(\rS)$ tends to 0, this realisation
constitutes an approximation of the space $\F$ and of its quantum noises.

\sspa{III.4}{Quantum Langevin equations}

In this article what we call quantum Langevin equation is actually a
restricted version of what is usually understood in the physical literature
(cf [G-Z]); by this we mean that we study here the so-called quantum
stochastic differential equations as defined by Hudson and
Parthasarathy and heavily studied by further authors ([H-P], [Fag]). 
This type of
quantum noise perturbation of the Schr\"odinger equation is exactly
the type of equation which we will get as the continuous limit of our
Hamiltonian description of repeated quantum interactions.
 \bigskip
Quantum stochastic differential equations are
operator-valued equations on $\rH_S\otimes\Phi$ of the form
$$
dU_t=\sum_{i,j=0}^NL^i_jU_t\, da^i_j(t),
$$
with initial condition $U_0=I$. The above equation has to be
understood as an integral equation 
$$
U_t=I+\int_0^t\sum_{i,j=0}^N L^i_jU_t\, da^i_j(t),
$$
the operators $L^i_j$ being bounded operators on $\rH_S$ alone which
are ampliated to $\rH_S\otimes\F$.

The main motivation and application of that kind of equation is that
it gives an account of the interaction of the small system $\rH_S$
with the bath $\F$ in terms of quantum noise perturbation of a
Schr\"odinger-like equation. Indeed, the first term of the equation 
$$
dU_t=L^0_0U_t\, dt+\ld
$$ 
describes the induced dynamics on the small system, all the other
terms are quantum noises terms. One of the main application of
these equations  is that they give 
explicit constructions of unitary dilations of semigroups of
completely positive maps of 
$\rB(\rH_S)$ (see [H-P] and also section VII of this article).  

Let us
here only recall one of  the main existence, uniqueness  and boundedness
theorem connected to quantum Langevin equations. The literature is
huge about those equations; we refer to [Par] for the result we
mention here.  In the following, by coherent vectors we mean elements
of the space
$\rE$ generated by the $u\otimes \varepsilon(f)$, with $u\in\rH_S$,
$f\in L^2(\Rp;\CC^n)$ and 
$$
[\varepsilon(f)](\s)=\prod_{(s,i)\in\s} f_i(s),
$$
the usual coherent vectors of the Fock space $\Phi$.

\th{5.}{\it If  all the operators $L^i_j$ are bounded on $\rH_S$
then the quantum
stochastic differential equation
$$
U_t=I+\sum_{i,j=0}^N\int_0^tL^i_jU_s\, da^i_j(s)
$$
admits a unique solution defined on the space of coherent vectors.

The solution $\pro U$ is made of unitary operators if and only if
 there exist on $\rH_S$, a self-adjoint operator $H$, operators
 $L_i$, $i=1,\ld,N$ and operators $S^i_j$, $i,j=1,\ld,N$ such that the matrix
 $(S^i_j)_{i,j=1,\ld,N}$ is unitary and the coefficients $L^i_j$ are of
 the form 
$$
\eq{
L^0_0&=-(iH+\frac 12\sum_{k=1}^NL^\ast_kL_k)\cr
L^0_j&=L_j \vphantom{\sum_{k=1}^N}\cr
L^i_0&=-\sum_{k=1}^NL^\ast_kS^k_i\cr
L^i_j&=S^i_j-\d_{ij}I.\cr
}
$$
}\qed

\sspa{III.5}{Convergence theorems}

We are finally able to state the main result of [AP1] which shows the
convergence of repeated interactions models to quantum stochastic
differential equations.

%operateurs et integrandes 

Let $\tau$ be a parameter in $\Rp$, which is thought of as
representing a small time interval. 
Let $U(\tau)$ be a unitary operator on $\rH_S\otimes\CC^{N+1}$, with coefficients
$U^i_j(\tau)$ as a matrix of operators on $\rH_S$ (this operator has
to be though of as  corresponding to the unitary operator $U$ of section
II). Let $V_k(\tau)$ be 
the associated repeated interaction operator: 
$$
V_{k+1}(\tau)=U_{k+1}(\tau)V_{k}(\tau)
$$
with the same notation as in section II.
In the following we will drop dependency in $\tau$ and write simply $U$, 
or $V_k$. Besides, we denote
$$
\e_{ij}=\frac 12(\d_{0i}+\d_{0j})
$$
for all $i,j$ in $\{0,\ld, N\}$. That is, for $i,j\geq 1$ 
$$
\e_{i0}=\e_{0j}=\frac 12,\ \ \e_{ij}=0, \ \ \e_{00}=1.
$$

Note that from now on we take the embeding of $\TF$ in $\F$ for
granted and we consider, without mentionning it,  all the repeated
quantum interactions to 
happen in $\TF(\tau)$, the subspace of $\F$ associated to the partition
$\rS=\{t_i=i\tau; i\in\NN\}$. The main result of [AP1] (Theorem 13 in
this reference) is the following.

\th{6.}{\it Assume that there exist bounded operators $L^i_j$, $i,j
\in \{0, \ld, n\}$ on
$\rH _S$ such that 
$$\lim_{\tau\rightarrow 0}
\frac{U^i_j(\tau)-\d_{ij}I}{\tau^{\e_{ij}}}=L^i_j
$$
for all $i,j=0,\ld,n$. Then, for almost all $t$ the operators
$V_{[t/\tau]}$ converge strongly, when $\tau\rightarrow 0$, to $V_t$,
the unitary  solution of the quantum stochastic 
differential equation
$$
dV_t=\sum_{i,j=0}^nL^i_jV_t\, da^i_j(t)
$$
with initial condition $V_0=I$.
}

\spa{IV.}{The G.N.S. representation of the heat bath}

In order to apply  Theorem 6 to our repeated interaction
model, we need the state of the photon to be a vector state (i.e. a
pure state) instead
of a density matrix. This is easily performed by considering the
so-called G.N.S. representation (or cyclic representation) of the
photon system.

This representation can be described in the following way. Consider the space
$\rH=\rL(\CC^{N+1})$ of endomorphisms of $\CC^{N+1}$. Consider a given
density matrix $\rho_\b$ on $\rH$, which is supposed to be in diagonal
form $\r_\b=\hbox{diag}(\b_0,\ldots,\b_N)$, where all the $\b_i$ are strictly
positive. The space $\rH=\rL(\CC^{N+1})$  is
made into a Hilbert space when equipped with the scalar product :
$$
\langle A,B\rangle=\tr(\r_\b\, A^*B),
$$
for all $A,B\in\rH$. The associated norm on $\rH$ is
denoted by $\norme\cd$. This Hilbert space is $(N+1)^2$-dimensional and 
we shall describe one of its orthonormal basis as follows. We denote
by $X^0_0$ the identity endomorphism. Then, for $i=1,\ld,N$, we put
$X^i_i$ to be the diagonal matrices with diagonal coefficients
$\{\l_i^1,\ld,\l_i^N\}$ such that 
$$
\langle X^i_i,X^j_j\rangle=\d_{ij}
$$
for all $i,j=0,1,\ld N$. Such a family clearly exists for its diagonal
elements are obtained
by extending the vector $(1,\ld,1)\in\CC^{N+1}$ into an orthonormal
basis of $\CC^{N+1}$ equiped with the scalar product
$$
\sum_{i=0}^N \b_i\,\bar x_iy_i.
$$
For $i\not =j\in\{0,\ld,N\}$ we put $X^i_j$ to be the element of $\rH$
given by 
$$
X^i_j={1\over\sqrt{\b_i}}\,a^i_j.
$$
It is then a straightforward computation to check that
$\{X^i_j;i,j=0,\ld N\}$ forms an orthonormal basis of $\rH$.
\bigskip
We now have the usual G.N.S. representation $\pi$ of $\rL(\CC^{N+1})$ into
$\rL(\rH)$ given by 
$$
\pi(A)B=AB,
$$
for all $A\in\rL(\CC^{N+1})$, $B\in\rH$. In the framework of that
representation, note that $X^0_0$ is then the vector state on $\rH$
which represents the state $\r_\b$ on $\CC^{N+1}$: indeed, for all
$A\in\rL(\CC^{n+1})$ we have
$$
\langle X^0_0, \pi(A)X^0_0\rangle=\tr (\r_\b\, A).
$$
That is, in the orthonormal basis we have choosen, $X^0_0$ is the only
important vector that could not be choosen to be different. The rest
of the choice for our orthonormal basis is just convenient for the
computations, but the final result does not depend on it.
\bigskip
Now, any operator $K$ on $\rH_S\otimes\CC^{N+1}$ is transformed by
$\pi$ into an operator on $\rH_S\otimes \rH$. That is, $\pi(K)$ is a
$(N+1)^2\times(N+1)^2$-matrix with coefficients $K^{i,j}_{k,l}$ in
$\rL(\rH_S)$. These coefficients are given by
$$
K^{i,j}_{k,l}=\tr_\rH(\r_\b\, (X^k_l)^*KX^i_j)
$$
where $\tr_\rH(H)$ denotes the partial trace
of $H$ along $\rH$, that is, this is the operator on $\rH_S$ given by
the sum of the diagonal coefficients of $H$  as a $\rL(\rH_S)$-valued
$(N+1)^2\times(N+1)^2$-matrix. 
\bigskip
When taking the continuous limit on the space $\rH_S\otimes\otimes_{\NNE}\rH$,
we end up into the space 
$$
\rH_S\otimes\bigotimes_{\Rp}\rH,
$$
that is, 
$$
\rH_S\otimes\G_s\left(L^2(\Rp;\CC^{(N+1)^2-1})\right).
$$
The associated quantum noises, following the same basis, are thus
denoted by $da^{i,j}_{k,l}(t)$, $i,j,k,l=0,\ld,N$.

\spa{V.}{The limit quantum Langevin equation}

We are now in conditions to apply 
Theorem 6. We consider the repeated interaction model described in
section II, with its associated operators $H$, $U_k$, $V_k$. Taking
the G.N.S. representation of all that we end up in the space
$\rH_S\otimes\otimes_{\NNE}\rH$, which we embbed inside a
continuous tensor product $\rH_S\otimes\otimes_{\Rp}\rH$, as explained
in section III.3. The main result of this article is then the following.

\th{7.}{\it In the continuous limit $\tau\rightarrow 0$, the repeated
interaction dynamics $V_{[t/\tau]}$  converges strongly on $\rH_S\otimes\Phi$, for  
all $t$, to the
(unitary) solution of the quantum Langevin equation
$$
\eq{
dU_t&=-\left[iH_S+i\sum_{i=0}^N\b_i\g_i\, I+{1\over
2}\sum_{i=1}^N(\b_0V_i^*V_i+\b_iV_iV_i^*)\right]U_t\, dt\cr
&\qq -i\sum_{i=1}^N\left[
\sqrt{\b_i}\,V_i\, U_t\, da^{i,0}_{0,0}(t)+\sqrt{\b_0}\, V_i^*\,U_t\,
da^{0,i}_{0,0}(t)\right.\cr
&\qq\left.+\sqrt{\b_i}\,V^*_i\,U_t\,
da^{0,0}_{i,0}(t)+\sqrt{\b_0}\,V_i\,U_t\, da^{0,0}_{0,i}(t)\right].&(8)\cr
}
$$
}
\prf
First, we represent the operators $H$ and $U$ on $\rH_S\otimes
\CC^{N+1}$ as $(N+1)\times(N+1)$-matrices with coefficients in
$\rL(\rH_S)$, following the orthonormal basis $\{e_0,e_1,\ld,
e_N\}$. We get
$$
H(\tau)=\left(
\matrix{
H_S+\g_0I&{1\over{\sqrt
\tau}}V_1^*&{1\over{\sqrt \tau}}V_2^*&\ldots&
{1\over{\sqrt \tau}}V_N^*\cr 
\ecarte{1\over{\sqrt
\tau}}V_1&H_S+\g_1 I&0&\ldots&0\cr 
\ecarte{1\over{\sqrt
\tau}}V_2&0&H_S+\g_2I&\ldots& 0\cr
\ecarte\vdots&\vdots&\cdots&\ddots&\vdots\cr
\ecarte{1\over{\sqrt
\tau}}V_N&0&0&\ldots&H_S+\g_NI\cr
}
\right).
$$

We need now to compute the associated unitary operator $U=e^{-i\tau H}$ in the
same framework. But as we wish to apply Theorem 6, we do not need to
wonder about the exact expression of $U$, but only the expansion of
its coefficients in powers of $\tau$ up to the pertinent orders given
by Theorem 6. We get that $U$ is represented by the matrix
\smallskip
$$
\left(
\matrix{
I-i\tau H_S-i\tau \g_0I&
-i\sqrt\tau\, V_1^*+o(\tau^{3/2})&
\ldots&
-i\sqrt\tau\, V_N^*+o(\tau^{3/2})\cr
-{1\over
2}\tau\sum_{i=1}^N V_i^*V_i+o(\tau^{2})&&&\cr
\cr
-i\sqrt\tau\, V_1+o(\tau^{3/2})&
I-i\tau H_S-i\tau\g_1I&
\ldots&
-{1\over 2}V_1V_N^*+o(\tau^2)\cr
&-{1\over 2}\tau V_1V_1^*+o(\tau^2)&&\cr
\cr
\vdots&\vdots&\ddots&\vdots\cr
\cr
-i\sqrt\tau\, V_N+o(\tau^{3/2})&
-{1\over 2}\tau V_NV_1^*+o(\tau^2)&
\ldots&
I-i\tau H_S-i\tau \g_NI\cr
&&&-{1\over 2}V_NV_N^*+o(\tau^2)\cr
}
\right).
$$
\smallskip
This is for the expression of $U$ as an operator on
$\rH_S\otimes\CC^{N+1}$. Now our aim is to compute the coefficients of
the matrix of $\pi(U)$ in 
the G.N.S. representation, that is, as a $(N+1)^2\times(N+1)^2$-matrix with
coefficients in $\rL(\rH_S)$.

As we already discussed in section IV, the  coefficients of $\pi(U)$ are obtained by
computing the quantities
$$
U^{i,j}_{k,l}=\langle X^k_l,\pi(U)X^i_j\rangle=\tr_{\rH}(\r_\b\,
(X^k_l)^\ast U 
X^i_j).
$$
In order to get $U^{0,0}_{0,0}$ we have to compute
$$
\tr_{\rH}(\r_\b\, X^0_0 U X^0_0)=\tr_{\rH}(\r_\b\, U).
$$
This gives, using $\sum_{i=0}^N\b_i=1$
$$
\eq{
U^{0,0}_{0,0}&= \b_0\left(I-i\tau H_S-i\tau \g_0I-{1\over
2}\tau\sum_{i=1}^NV_i^*V_i+o(\tau^{2})\right)+\cr
&\ \ \ +\b_1\left(I-i\tau H_S-i\tau
\g_1I -{1\over 2}\tau V_1V_1^*+o(\tau^2)\right)+\ld\cr
&\ \ \ +\b_N\left(I-i\tau H_S-i\tau \g_NI-{1\over
2}\tau V_NV_N^*+o(\tau^2)\right)\cr
U^{0,0}_{0,0}&=I-i\tau H_S-i\tau\sum_{i=1}^N\b_i \g_i\,I-{1\over
2}\tau\sum_ {i=0}^N(\b_0V_i^*V_i+\b_iV_iV_i^*)+o(\tau^2).&(9)\cr 
}
$$
We now compute the $U^{i,j}_{0,0}$ terms, for $i\not=j$. That is, we compute
$$
U^{i,j}_{0,0}=\tr_{\rH}(\r_\b\, U X^i_j).
$$
This trace is equal to 
$$
{1\over\sqrt{\b_i}}\langle e_i,\r_\b\, Ue_j\rangle=
{\sqrt{\b_i}}\,\langle e_i,Ue_j\rangle.
$$
Then two distinct cases appear. If $j=0$ we have
$$
Ue_0=\left(\matrix{
I-i\tau H_S-i\tau \g_0I
-{1\over
2}\tau\sum_{i=1}^NV_i^*V_i+o(\tau^{2})
\cr
\ecarte -i\sqrt\tau\, V_1
+o(\tau^{3/2})
\cr
\ecarte\vdots
\cr
\ecarte-i\sqrt\tau\, V_N
+o(\tau^{3/2})\cr
}
\right)
$$
and thus 
$$
U^{i,0}_{0,0}=-i\sqrt\tau {\sqrt{\b_i}}\,V_i+o(\tau^{3/2}).\eqno{(10)}
$$
But when $j\not=0$ we have
$$
Ue_j=\left(\matrix{
-i\sqrt\tau \,V_j^*+o(\tau^{3/2})
\cr
\ecarte-{1\over 2}\tau V_1V_j^*+o(\tau^2)\cr
\ecarte\vdots\cr
\ecarte I-i\tau H_S-i\tau \g_jI
-{1\over
2}\tau V_jV_j^*+o(\tau^{2})
\cr
\ecarte\vdots\cr
\ecarte-{1\over 2}\tau V_NV_j^*+o(\tau^2)\cr
}
\right).
$$
Now, if $i=0$ we get
$$
U^{0,j}_{0,0}=-i{\sqrt{\b_0}}\sqrt\tau\, V_j^*+o(\tau^{3/2}),\eqno{(11)}
$$
if $i\not=0$ and $i\not =j$ we get
$$
U^{i,j}_{0,0}=-{1\over 2}{\sqrt{\b_i}}\,\tau\,
V_iV_j^*+o(\tau^2). \eqno{(12)}
$$
A similar computation gives
$$
U^{0,0}_{0,j}=-i\sqrt\tau\sqrt{\b_0\,}V_j+o(\tau^{3/2}),\eqno{(13)}
$$
$$
U^{0,0}_{i,0}=-i\sqrt{\b_i}\sqrt\tau\, V_i^*+o(\tau^{3/2})\eqno{(14)}
$$
and, still for $i\not =j$ and $i,j\not=0$
$$
U^{0,0}_{i,j}=-{1\over 2}\sqrt{\b_i}\,\tau\, V_jV_i^*+o(\tau^2).\eqno{(15)}
$$
Now, consider $i\not=j$ and $k\not=l$, we have
$$
\eq{
U^{i,j}_{k,l}&=\tr_\rH(\r_\b\, (X^k_l)^*UX^i_j)=\sum_p\langle X^k_l\r_\b\,
e_p,UX^i_j \,e_p\rangle\cr
&=\sum_p \b_p{1\over \sqrt{\b_k\b_i}}\langle
a^k_le_p,Ua^i_je_p\rangle\cr
&= \d_{ik}\langle e_l,Ue_j\rangle.
}
$$ 
That is, if $l=0$ and $j\not =0$ 
$$
U^{i,j}_{k,0}=\d_{ik}(-i\sqrt\tau V_j^*+o(\tau^{3/2}))\eqno{(16)}
$$
and for $l=j=0$
$$
U^{i,0}_{k,0}=\d_{ik}(I-i\tau H_S-i\tau\g_0
I-{1\over2}\sum_{i=1}^NV_i^*V_i+o(\tau^2)).
\eqno{(17)}
$$
If $l\not=0$ and $l\not=j$ with $j\not =0$ we have
$$
U^{i,j}_{k,l}=\d_{ik}(-{1\over 2}\tau\, V_lV_j^*+o(\tau^2)),\eqno{(18)}
$$
if $l=j\not=0$
$$
U^{i,j}_{k,j}=\d_{ik}(I-i\tau H_S-i\tau\g_l\,I-{1\over 2}\tau\,
V_lV_l^*+o(\tau^2))\eqno{(19)} 
$$
and finally if $l\not=0$ and $j=0$ then
$$
U^{i,0}_{k,l}=\d_{ik}(-i\sqrt\tau V_l+o(\tau^{3/2})).\eqno{(20)}
$$
We now study the $U^{i,i}_{j,j}$ terms. Recall that the 
$X^i_i$ are diagonal matrices whose coefficients $(\l^0_i,\ldots
\l^N_i)$, $i=0,\ld,N$ form an orthonormal basis of 
$\CC^{N+1}$ equipped with the scalar product $\sum_i\b_i
\bar{x_i}y_i$. We get
$$
\eq{
U^{i,i}_{0,0}&=\tr(\r_\b\, U X^i_i)=\sum_{k=0}^N\langle e_k,\r_\b\, U X^i_i
e_k\rangle\cr
&=\sum_{k=0}^N \b_k \l^k_i \langle e_k,Ue_k\rangle\cr
&=\sum_{k=1}^N \b_k\l^k_i (I-i\tau H_S-i\tau \g_kI-{1\over 2}\tau\,
V_kV_k^*+o(\tau^{2}))\cr
&\qq+\b_0\l^0_i(I-i\tau H_S-i\tau\g_0
I-{1\over2}\sum_{i=1}^NV_i^*V_i+o(\tau^2))\cr
&=\sum_{k=1}^N \b_k\l^k_i (-i\tau \g_kI-{1\over 2}\tau\,
V_kV_k^*+o(\tau^{2}))\cr
&\qq+\b_0\l^0_i(-i\tau\g_0
I-{1\over2}\sum_{i=1}^NV_i^*V_i+o(\tau^2))&(21)\cr
}
$$
where we have used $\sum_k \b_k\l^k_i=0$.
In the same way,
$$
\eq{
U^{0,0}_{i,i}&=\sum_{k=1}^N \b_k\overline{\l^k_i} (-i\tau
\g_k\,I-{1\over 2}\tau\,
V_kV_k^*+o(\tau^{2}))\cr
&\qq+\b_0\overline{\l^0_i}(-i\tau\g_0
I-{1\over2}\sum_{i=1}^NV_i^*V_i+o(\tau^2)).&(22)\cr
}
$$
Finally, using $\sum_{k=0}^N \b_k\,\overline{\l^k_j}\l^k_i=\d_{ij}$, we obtain 
$$
\eq{
U^{i,i}_{j,j}&=\sum_{k=0}^N \b_k \,\overline{\l^k_j}\l^k_i \langle
e_k,Ue_k\rangle\cr
&=\d_{ij}(I-i\tau H_S)+\sum_{k=1}^N \b_k\,\overline{\l^k_j}\l^k_i (-i\tau
\g_kI-{1\over 2}\tau\,
V_kV_k^*+o(\tau^{2}))\cr
&\qq+\b_0\overline{\l^0_j}\l^0_i(-i\tau\g_0-{1\over 2}\tau\sum_{j=1}^NV_j^*V_j+o(\tau^2)).&(23)\cr
}
$$
The last terms to be considered are those of type $U^{i,i}_{k,l}$ (and
conversely $U^{k,l}_{i,i}$),
with $k\not=l$ and $i\not =0$. We
have
$$
\eq{
U^{i,i}_{k,l}&=\tr(\r_\b\,(X^k_l)^* U X^i_i)=\sum_p\langle
e_p,\r_\b\,(X^k_l)^* U X^i_i 
e_p\rangle\cr
&=\sum_p \b_p\, \l^p_i \langle X^k_l e_p,Ue_p\rangle\cr
&=\b_k\,\l^k_i{1\over \sqrt{\b_k}}\langle e_l, Ue_k\rangle\cr
}
$$
which gives 
$$
\sqrt{\b_0}\,\l^0_i(-i\sqrt \tau V_l+o(\tau^{3/2}))\eqno{(24)}
$$
or 
$$
\sqrt{\b_k}\l^k_i(-{1\over 2} \tau V_lV_k^*+o(\tau^{2}))\eqno{(25)}
$$
depending on $k=0$ or not. The case of $U_{i,i}^{k,l}$ is similar and
needs not be explicited for anyway it will not contribute to the
continuous limit.
\bigskip
We can now apply Theorem 6. 
Following the rules of Theorem 6, we need to check that there exists
bounded operators $L^{i,j}_{k,l}$ on $\rH_S$ such that
$$
s-\lim_{\tau\rightarrow 0} {{U^{i,j}_{k,l}-\d_{(i,j), (k,l)}\,I}\over
\tau^{\e^{i,j}_{k,l}}}=L^{i,j}_{k,l}\,,
$$
where $\e^{0,0}_{0,0}=1$, $\e^{0,0}_{k,l}=\e_{0,0}^{k,l}=1/2$ and the
others $\e^{i,j}_{k,l}$ are equal to 0.

Equality (9) shows that 
$$
L^{0,0}_{0,0}=-iH_S-i\sum_{i=1}^N\b_i\g_iI-{1\over
2}\sum_i(\b_0V_i^*V_i+\b_iV_iV_i^*).
$$
By (10) we have
$$
L^{i,0}_{0,0}=-i\sqrt{\b_i}V_i
$$
and in the same way 
$$
\eq{
L^{0,i}_{0,0}&=-i\sqrt{\b_0}V_i^\ast\cr
L^{0,0}_{i,0}&-i\sqrt{\b_i}V_i^*\cr
L^{0,0}_{0,i}&=-i\sqrt{\b_0}V_i\cr
}
$$
by (11), (14) and (13) respectively.

The other terms $U^{i,j}_{0,0}$ and $U^{0,0}_{i,j}$ appear to be of order
$\tau$ in (12) and (15), while only their $\sqrt\tau$ part contributes
to the limit. As a consequence $L^{i,j}_{0,0}=L^{0,0}_{i,j}=0$.

Terms of the form $U^{i,j}_{k,l}$ (equations (16) to (20)) contribute
in the limit via the order 1 terms in
$U^{i,j}_{k,l}-\d_{(i,j),(k,l)}\, I$, that is 0 in anycase (the $I$
term in (17) and (19) does not contribute as it indeed appears only when $(i,j)=(k,l)$.

The same holds for $U^{i,i}_{0,0}$ and $U^{0,0}_{i,i}$ which gives
$L^{i,i}_{0,0}=L^{0,0}_{i,i}=0$ (equations (21) and (22)).

The terms $U^{i,i}_{j,j}$ contribute in the limit via the order 1
terms of  $U^{i,i}_{j,j}-\d_{ij} I$. Following (23) we get a null
contribution in all cases.

Finally, equality (24) and (25) show that the last  coefficients
also vanish in the limit.

This exactly gives the announced quantum Langevin equation.\qed
\bigskip
{\bf Remark:} One can only be impressed (at least that was the case of
the authors when performing the computations) by the kind of
``mathematical miracle'' occuring here: all the $(N+1)^2$ terms fit
perfectly in the type of conditions of Theorem 6. The number of
cancellation one may hope for happens exactly. 
\bigskip
Equation (8) takes a much more useful form if one
regroups correctly the different terms. Indeed, put 
$$
A^0_i(t)=\sqrt{{{\b_0}\over{\b_0-\b_i}}}\,
a^{0,0}_{0,i}(t)+\sqrt{{{\b_i}\over{\b_0-\b_i}}}\, a^{i,0}_{0,0}(t) \eqno{(26)}
$$
$$
A^i_0(t)=\sqrt{{{\b_0}\over{\b_0-\b_i}}}\,a^{0,i}_{0,0}(t)+
\sqrt{{{\b_i}\over{\b_0-\b_i}}}\,a^{0,0}_{i,0}(t)\eqno{(27)}
$$
and
$$
W_i=-i\sqrt{\b_0-\b_i}\,V_i
$$
then the  equation (8) simply writes
$$
\eq{
dU_t&=-\left[iH_S+i\sum_{i=0}^N\b_i\g_i\,I+{1\over
2}\sum_{i=1}^N\left({{\b_0}\over{\b_0-\b_i}}W_i^*W_i+{{\b_i}
\over{\b_0-\b_i}}W_iW_i^*\right)\right]U_t\, 
dt\cr 
&\ \ \ +\sum_i\left(W_i U_t\, dA^0_i(t)-W^*_i U_t\, dA^i_0(t)\right).&(28)\cr
}
$$

\spa{VI.}{Thermal quantum noises and their properties}

In this section we concentrate on the particular quantum noises (26)
and (27) that appeared above. We shall show that they are natural candidates
for being qualified as ``thermal quantum noises''. We also show that
the form of equation (25) is the generic one for unitary solutions, in
the thermal case.
\bigskip
The situation that has appeared in the previous section can be summarized
as follows. 

We consider the quantities $\b$, $\g_i$ and thus $\b_i$ as being fixed.

First of all,  there is no need  to consider a Fock space over
$L^2(\Rp; \CC^{(N+1)^2-1})$ anymore, for most of the quantum noises
$a^{i,j}_{k,l}(t)$ do not play any role in equation (8). More economical is to
consider a double Fock space:
$$
\wt\F=\G_s(L^2(\Rp,\CC^{N}))\otimes \G_s(L^2(\Rp,\CC^{N})).
$$
Each of the copies of the Fock space accomodates the  quantum noises
$$
a^i_j(t)\otimes I\ \ \hbox{and}\ \ I\otimes a^i_j(t)
$$
respectively, which we shall denote more simply by
$$
a^i_j(t)\ \ \hbox{and}\ \ b^i_j(t)
$$
respectively.

Form the operator processes
$$
A^0_i(t)=\sqrt{{{\b_0}\over{\b_0-\b_i}}}\,
a^{0}_{i}(t)+\sqrt{{{\b_i}\over{\b_0-\b_i}}}\, b^{i}_{0}(t) \eqno{(28)}
$$
$$
A^i_0(t)=\sqrt{{{\b_0}\over{\b_0-\b_i}}}\,a^{i}_{0}(t)+
\sqrt{{{\b_i}\over{\b_0-\b_i}}}\,b^{0}_{i}(t).\eqno{(29)}
$$
For every $f\in L^2(\Rp;\CC^n)$ with coordinates $(f_i)$ in the basis
$\{e_1,\ld,e_n\}$ put
$$
A^*(f)=\sum_{i=1}^n\int_{\Rp} f_i(t)\, dA^0_i(t)
$$
and 
$$
A(f)=\sum_{i=1}^n\int_{\Rp} \overline{f_i(t)}\, dA^i_0(t).
$$
\prp{8.}{\it The operators $A(f),A^*(g)$ form a non-Fock representation of the
CCR algebra over $(L^2(\Rp;\CC^N))$.}
\prf
The operators $A(f)$ and $A^*(g)$ have similar properties as the usual
quantum noises. In particular,
they admit a quantum stochastic integration theory, which is
completely identical to the usual one. This does not need to be
developed here. We shall only prove that the quantum Ito formula
(7) is now driven by the rules:
$$
dA^i_0(t)\, dA^0_i(t)={{\b_0}\over{\b_0-\b_i}}\, dt
$$
and 
$$
dA_i^0(t)\, dA_0^i(t)={{\b_i}\over{\b_0-\b_i}}\, dt.
$$
Indeed, we have
$$
\eq{
&dA^i_0(t)\, dA^0_i(t)=\cr
&=\left(\sqrt{{{\b_0}\over{\b_0-\b_i}}}\,da^{i}_{0}(t)+
\sqrt{{{\b_i}\over{\b_0-\b_i}}}\,db^{0}_{i}(t)\right)
\left(\sqrt{{{\b_0}\over{\b_0-\b_i}}}\,  
da^{0}_{i}(t)+\sqrt{{{\b_i}\over{\b_0-\b_i}}}\, db^{i}_{0}(t)\right)\cr
&={{\b_0}\over{\b_0-\b_i}}\, da^0_0(t)={{\b_0}\over{\b_0-\b_i}}\,
dt\cr
}
$$
and
$$
\eq{
&dA_i^0(t)\, dA_0^i(t)=\cr
&=
\left(\sqrt{{{\b_0}\over{\b_0-\b_i}}}\,  
da^{0}_{i}(t)+\sqrt{{{\b_i}\over{\b_0-\b_i}}}\, db^{i}_{0}(t)\right)
\left(\sqrt{{{\b_0}\over{\b_0-\b_i}}}\,da^{i}_{0}(t)+ 
\sqrt{{{\b_i}\over{\b_0-\b_i}}}\,db^{0}_{i}(t)\right)\cr
&={{\b_i}\over{\b_0-\b_i}}\, db^0_0(t)={{\b_i}\over{\b_0-\b_i}}\,
dt\,.\cr
}
$$
By the quantum Ito formula we get
$$
\eq{
[A(f),A^*(g)]&=\sum_{i=1}^N\int_{\Rp} 
\overline{f_i(t)}\, g_i(t)\, \left({{\b_0-\b_i}\over{\b_0-\b_i}}\right)dt\,
I\cr
&=\langle f,g\rangle I.\cr
}
$$
In other words,  the operators $A(f),A^*(g)$ form a representation of the
CCR algebra over $(L^2(\Rp;\CC^N))$.  But this clearly a non-Fock one
for the creation and annihilation operator attached to this
representation do not generate the whole creation and annihilation
operators of the underlying (double) Fock space.\qed
\bigskip
Now, let us form the associated Weyl operators
$$
W(f)=\exp\left({{A(f)+A^*(f)}\over{\sqrt 2}}\right).
$$
We wish to compute the statistics of $W(f)$ in the vaccum state
$\O$. For this purpose, we use the following notation. If $H$ is any
operator on $\CC^{N+1}$, then it acts on $L^2(\Rp;\CC^N)$ by
$$
[Hf](s)=\l_0+\sum_{i=1}^N\l_if_i(s).
$$
This has to be understood as follows: in general $\l_0$ is chosen to
be equal to 0, thus $H$ acts on $L^2(\Rp;\CC^N)$ as a multiplication
operator. In our case it is the multiplication by a constant (vector).

\th{9.}{\it We have
$$
\langle \O,W(f)\,\O\rangle=\exp\left(-{1\over
4}\langle f,\coth(\b {H_R\over 2})f\rangle\right)
$$
for all $f\in L^2(\Rp;\CC^N)$.}
\prf
We have
$$
\di{
A(f)+A^*(f)=\sum_{i=1}^N\int_0^\infty\left(\sqrt{{{\b_0}\over{\b_0-\b_i}}}\,
f_i(s)\, da^{0}_{i}(s)+\sqrt{{{\b_i}\over{\b_0-\b_i}}}\, f_i(s)\,
db^{i}_{0}(s)\right.+\hf\cr
\hf+\left.\sqrt{{{\b_0}\over{\b_0-\b_i}}}\,\overline{f_i(s)}\, da^{i}_{0}(s)+
\sqrt{{{\b_i}\over{\b_0-\b_i}}}\,\overline{f_i(s)}\, db^{0}_{i}(s)\right).\cr
}
$$
If we put 
$$
a(f)=\sum_{i=1}^N\int_0^\infty \left(f_i(s)\,
da^0_i(s)+\overline{f_i(s)}\, da^i_0(s)\right)
$$
and 
$$
b(f)=\sum_{i=1}^N\int_0^\infty \left(f_i(s)\,
db^0_i(s)+\overline{f_i(s)}\, db^i_0(s)\right)
$$
then the above expression shows that
$$
A(f)+A^*(f)=a(\wt f)+b(\wh f)
$$
where
$$
\eq{
\wt f_i(s)&=\sqrt{{{\b_0}\over{\b_0-\b_i}}}\, f_i(s)\cr
\wh f_i(s)&=\sqrt{{{\b_i}\over{\b_0-\b_i}}}\, \overline{f_i}(s).\cr
}
$$
Denote by $W_a$ and $W_b$ the usual Weyl operators associated to the
noises $a$ and $b$ respectively. We have clearly shown that
$$
W(f)=W_a(\wt f)\otimes W_b(\wh f).
$$
As a consequence, using usual computations on the Weyl operators
$$
\eq{
\langle\O,W(f)\O\rangle&=\exp\left(-{1\over
4}\left(\normca{\wt f}+\normca{\wh f}\right)\right)\cr
&=\exp\left(-{1\over
4}\sum_{i=1}^n\left({{\b_0}\over{\b_0-\b_i}}+{{\b_i}\over{\b_0-\b_i}}\right)
\norme{f_i}^2 \right)\cr
&=\exp\left(-{1\over
4}\sum_{i=1}^n\left(\coth(\b{{\g_i}\over 2})\norme{f_i}^2
\right)\right)\cr
&=\exp\left(-{1\over
4}\langle f,\coth(\b {H_R\over 2})f\rangle\right).&\qed\cr
}
$$
\bigskip
We recover an analogue of the usual K.M.S. state statistics for a free Boson gas at
thermal equilibrium. Let us discuss that point more
precisely. Usually, the Hamiltonian model for a
quantum heat bath is as follows. We are given a function $\o(s)$ (in
Fourier representation actually, by this does not matter much here)
and the Hamiltonian of the heat bath, on the Fock space
$\Gamma_s(L^2(\Rp;\CC))$ is the differential second quantization
operator $d\Gamma(\o)$ associated to the multiplication by $\o$. 

In our discrete model, if we take the typical interacting system
$\rH_R$ to be $\CC^2$ but with a Hamiltonian depending on the number
of the copy:
$$
H_R(k)=\left(\matrix{0&0\cr 0&\o(k)}\right)
$$
then in the continuous limit, the corresponding Hamiltonian on the
Fock space  $\Gamma_s(L^2(\Rp;\CC))$ is indeed  $d\Gamma(\o)$ (under
some continuity assumption on $\o$, cf [AP2]). 

In the case we have described here, the situation is made a little
more complicated by the fact that we considered a chain of $\CC^{N+1}$
instead of $\CC^2$, but a lot easier by taking a constant Hamiltonian
$H_R$. The time-dependent case stays to be explored, no doubt it will
give rise to the usual free Bose gaz statistics.
\bigskip
{\bf Remark :} The parameter $\b$ used here is supposed to be the
inverse of the temperature of the heat bath (more exactly $1/kT$). If
we make the temperature go to 0, that is, $\b$ goes to $+\infty$, then
$$
{{\b_0}\over{\b_0-\b_i}}={1\over{1-e^{-\b(\g_i-\g_0)}}}
$$
converges to 1, for $\g_i-\g_0> 0$ by hypothesis, and 
$$
{{\b_i}\over{\b_0-\b_i}}={{e^{-\b\g_i}}\over{1-e^{-\b(\g_i-\g_0)}}}
$$
converges to 0. This makes all the noises $b^i_j$ being useless and
$A(g)=a(g)$, for all $g$. We recover the usual quantum noises, the
usual Weyl operators. This means that the usual quantum noises are the
0 temperature ones.
\bigskip
We now describe which kind of quantum Langevin equation, driven by
those thermal quantum noises gives rise to a unitary evolution.

\th{10.}{\it 
A Langevin equation of the form
$$
dU_t=K^0_0U_t\, dt+\sum_{i=0}^n \left(K^0_iU_t\, dA^0_i(t)+K_0^iU_t\,
dA_0^i(t)\right),\qq U_0=I,
$$
where the coefficients $K^i_j$ are all bounded operators on
$\rH_S$, always admit a unique solution on the set of coherent
vectors. The solution is unitary if and only if  it is of the form
$$
\di{
dU_t=\left(-iH-{1\over 2}\sum_{i=1}^n\left(
{{\b_0}\over{\b_0-\b_i}}W_i^*W_i+{{\b_i}\over{\b_0-\b_i}}W_iW_i^*\right)\right)U_t\,
dt+\hf\cr
\hf+\sum_{i=0}^n \left(W_iU_t\, dA^0_i(t)-W_i^*U_t\, 
dA_0^i(t)\right)\cr
}
$$
for some bounded operators $W_i$, $i=1,\ld ,n$, on $\rH_S$ and a self-adjoint bounded
operator $H$ on $\rH_S$.}
\prf
The existence and uniqueness result is a simple consequence of the one
quoted in Theorem 5.

For characterizing the unitarity we use algebraical (formal) computations, which are
the same as for the proof of Theorem 5. The analytical part of the
proof is totaly identical to the one of Theorem 5. There is no need to
develop it here.

Our equation is of the general form
$$
dU_t=K^0_0U_t\, dt+\sum_{i=0}^n \left(K^0_iU_t\, dA^0_i(t)+K_0^iU_t\,
dA_0^i(t)\right)
$$
We thus
also have
$$
dU_t^*=U^*_t(K^0_0)^*\, dt+\sum_{i=0}^n \left(U_t^*(K^0_i)^*\,
dA_0^i(t)+U^*_t(K_0^i)^*\, 
dA^0_i(t)\right).
$$
By the quantum Ito formula we get
$$
\eq{
d(U_t^*&U_t)=(dU^*_t)U_t+U^*_t\, dU_t+dU^*_t\, dU_t\cr
&=U^*_t(K^0_0)^*U_t\, dt+\sum_{i=0}^n \left(U_t^*(K^0_i)^*U_t\, dA_0^i(t)+U^*_t(K_0^i)^*U_t\,
dA^0_i(t)\right)+\cr
&\ \ \ +U_t^*K^0_0U_t\, dt+\sum_{i=0}^n \left(U_t^*K^0_iU_t\,
dA^0_i(t)+U_t^*K_0^iU_t\, 
dA_0^i(t)\right)+\cr
&\ \ \ +{{\b_0}\over{\b_0-\b_i}}\sum_{i=1}^n U_t^*(K^0_i)^*K^0_iU_t\,
dt+{{\b_i}\over{\b_0-\b_i}}\sum_{i=1}^n U_t^*(K_0^i)^*K_0^iU_t\, dt\cr 
&=U^*_t\left((K^0_0)^*+K^0_0+\sum_{i=1}^n\left({{\b_0}\over{\b_0-\b_i}}(K^0_i)^*K^0_i
+{{\b_i}\over{\b_0-\b_i}}(K_0^i)^*K_0^i\right)\right)
U_t\, dt \cr
&\ \ \ +\sum_{i=0}^n \left(U_t^*\left((K^0_i)^*+K_0^i\right)U_t\right)\, dA_0^i(t)
+\sum_{i=0}^n \left(U_t^*\left((K_0^i)^*+K^0_i\right)U_t\right)\,
dA^0_i(t).\cr
}
$$
By a similar computation we obtain
$$
\di{
d(U_tU_t^*)=\hf\cr
=\left(U_tU^*_t(K^0_0)^*+K^0_0U_tU^*_t+\sum_{i=1}^n\left(
{{\b_0}\over{\b_0-\b_i}}K^i_0U_tU^*_t(K^i_0)^* +\right.\right.\hf\cr
\hf+\left.\left.{{\b_i}\over{\b_0-\b_i}}K_i^0U_tU^*_t(K_i^0)^*\right)\right)
\, dt \cr
\hf+\sum_{i=0}^n \left(U_tU^*_t(K^0_i)^*+K_0^iU_tU^*_t\right)\, dA_0^i(t)+ +\sum_{i=0}^n \left(U_tU^*_t(K_0^i)^*+K^0_iU_tU^*_t\right)\,
dA^0_i(t).\cr
}
$$
Asking both to be equal to 0 for every $t$ is equivalent to the following conditions :
$$
\di{
K^i_0=-(K^0_i)^*\hf\cr
(K^0_0)^*+K^0_0+\sum_{i=1}^n\left(
{{\b_0}\over{\b_0-\b_i}}(K^0_i)^*K^0_i+{{\b_i}\over{\b_0-\b_i}}K^0_i(K^0_i)^*\right)=0.\hf\cr
}
$$
Put 
$$
K=K^0_0+{1\over 2}\sum_{i=1}^n\left(
{{\b_0}\over{\b_0-\b_i}}(K^0_i)^*K^0_i+{{\b_i}\over{\b_0-\b_i}}K^0_i(K^0_i)^*\right),
$$
the last condition above exactly says
$$
K=-K^*.
$$
We thus obtain the announced characterization.
\qed
\bigskip
The attentive reader has noticed that this is exactly the form of
equation (28)!

\spa{VII.}{The Lindblad generator}

Going back to the usual Langevin equations of Theorem 5, let us recall
a very important theorem, 
which is the main point in using quantum Langevin equations in order to
dilate quantum dynamical semigroups.

\th{11.}{\it Consider the unitary solution $\pro U$ of the quantum
Langevin equation
$$
dU_t=-(iH+\frac 12\sum_{k=1}^NL^\ast_kL_k)U_t\, dt+\sum_{i=0}^NW_i
U_t\, da^0_i(t)-\sum_{i=0}^n W^*_k
U_t\, da^i_0(t).
$$
For any bounded operator $X$ on $\rH_S$, the application
$$
t\mapsto P_t(X)=\langle \O,U^*_t(X\otimes I)U_t\O\rangle
$$
is a semigroup of completely positive maps whose Lindblad generator is
$$
\rL(X)=i[H,X]-{1\over 2}\sum_{i=1}^N (W_i^*W_i
X+XW^*_iW_i-2W^*_iXW_i).
$$
\qed
}

In our thermal case we have the following the form for the Lindblad
generator.

\th{12.}{\it Consider the unitary solution $\pro U$ of the thermal quantum
Langevin equation
$$
\di{
dU_t=\left(-iH-{1\over 2}\sum_{i=1}^N\left(
{{\b_0}\over{\b_0-\b_i}}W_i^*W_i+{{\b_i}\over{\b_0-\b_i}}W_iW_i^*\right)\right)U_t\,
dt+\hf\cr
\hf+\sum_{i=0}^N \left(W_iU_t\, dA^0_i(t)-W_i^*U_t\, 
dA_0^i(t)\right).\cr
}
$$
For any bounded operator $X$ on $\rH_S$, the application
$$
t\mapsto P_t(X)=\langle \O,U^*_t(X\otimes I)U_t\O\rangle
$$
is a semigroup of completely positive maps whose Lindblad generator is
$$
\eq{
\rL(X)=i[H,X]&-{1\over2}\sum_{i=1}^N
{{\b_0}\over{\b_0-\b_i}}\left(W^*_iW_iX+XW^*_iW_i-2W^*_iXW_i\right)\cr
&-{1\over2}\sum_{i=1}^N
{{\b_i}\over{\b_0-\b_i}}\left(W_iW^*_iX+XW_iW^*_i-2W_iXW^*_i\right).
\cr
}
$$
}
\prf
The basic computation is the same as for Theorem 11: 

-- By the ``thermal quantum
Ito formula'' (see proof of Proposition 8) one computes
$$
d(U^*_t(X\otimes I)U_t)=(dU^*_t)(X\otimes I) U_t+U^*_t(X\otimes
I)(dU_t)+(dU^*_t)(X\otimes I) (dU_t);
$$

-- The only contributing term when averaging over the vacuum state is
   the coefficient of $dt$, that is, we get the equation
$$
d\langle \O,U^*_t(X\otimes I)U_t\O\rangle=\langle
\O,U^*_t(\rL(X)\otimes I)U_t\O\rangle\, dt.
$$
The solution is clearly a semigroup with generator $\rL$. \qed

Let us write down in a corollary, the Lindblad generator in
perspective with the initial Hamiltonian.

\co{13}{\it If the repeated interaction model is having the following
total Hamiltonian:
$$
H=H_S\otimes I+I\otimes H_{R}+{1\over{\sqrt\tau}}\sum_{i=1}^N
\left(V_i\otimes a^0_i+V_i^*\otimes 
a^i_0\right)
$$
then the associated Lindblad generator in the continuous limit is
$$
\eq{
\rL(X)=i[H_S,X]&-{1\over2}\sum_{i=1}^N
{{\b_0}}\left(V^*_iV_iX+XV^*_iV_i-2V^*_iXV_i\right)\cr
&-{1\over2}\sum_{i=1}^N
{{\b_i}}\left(V_iV^*_iX+XV_iV^*_i-2V_iXV^*_i\right).
\cr
}
$$
}

\spa{VIII.}{Thermalization}

In this section we answer a very natural question in this
context. Consider a given quantum system $\rH_S$ with a given
Hamiltonian $H_S$. Is there a natural Lindblad generator $\rL$ (in the
Schr\"odinger picture) on $\rH_S$
which admits as a unique invariant state, the state
$$
\rho_\b={1\over Z_\beta}\,e^{-\beta H_S}
$$
and which possesses the property of return to equilibrium for this
state? By ``return to equilibrium'' we mean the following: for every
inital state $\rho_0$, the evolution $e^{t\rL}(\rho_0)$ converges to
the state $\rho_\b$ in the $*$-weak sense, that is, 
$$
\lim_{t\rightarrow +\infty}\tr(e^{t\rL}(\rho_0)\, X)=\tr(\rho_\beta\,
X)
$$
for all observable $X$.

We shall prove in this section that the answer to the above question is positive, at
least if $\rH_S$ is 
finite-dimensional. For this purpose we recall a famous result by
Frigerio and Veri [F-V], in a slightly extended form due to Fagnola
and Rebolledo [F-R].

\th{14}{\it Let 
$$
\rL(\rho)=-i[H,\rho]-{1\over 2}\sum_{i=1}^n\left(L^*_iL_i\rho+\rho
L^*_iL_i-2L_i\rho L_i^*\right)
$$
be a Lindblad generator (in Schr\"odinger picture). If the commutants
$$
\left\{H,L_i,L_i^*;i=1,\ldots
n\right\}'\qq\hbox{and}\qq\left\{L_i,L_i^*;i=1,\ldots n\right\}'
$$
coincide then the associated dynamics possesses the property of return
to equilibrium.
}

\smallskip
We consider $\rH_S$ a $N+1$-dimensional Hilbert space, with
Hamiltonian (in diagonal form)
$$
H_S=\left(\matrix{\l_0&0&\ldots&0\cr0&\l_1&\ldots&0\cr\vdots&&\ddots&\vdots\cr
0&\ldots&\ldots&\l_N}\right)\,.
$$
Consider the Gibbs state $\rho_\b=(1/Z_\beta)\,e^{-\beta H_S}$, it is also
of diagonal form with diagonal elements  
denoted by $\b_0,\b_1,\ldots,\b_N$.

We put the system $\rH_S$ in repeated quantum interaction with a chain
of copies of $\CC^{N+1}$ with the total Hamiltonian
$$
H=H_S\otimes I+I\otimes H_{R}+{1\over{\sqrt\tau}}\sum_{i=1}^N
\left(V_i\otimes a^0_i+V_i^*\otimes 
a^i_0\right), 
$$
where $V_i$ is the matrix
$$
V_i=\left(\matrix{0&\ldots&1&\ldots&0\cr 0&0&\ldots&&0\cr
\vdots&&&&\vdots\cr
0&&0&\ldots&0}\right)
$$
with the 1 being at the $i$-th row and where $H_R=H_S$.

Note that this means that we put the system $\rH_R$ in repeated
quantum interaction with a chain of copies of ... itself but in the
desired state. This is somehow very natural!

\th{15}{\it In the continuous limit the above repeated interaction
model admits the following Lindblab generator in the Schr\"odinger
picture:
$$
\eq{
\rL(\rho)=-i[H_S,\rho]&-{1\over2}\sum_{i=1}^N
{{\b_0}}\left(V^*_iV_i\rho+\rho V^*_iV_i-2V_i\rho V^*_i\right)\cr
&-{1\over2}\sum_{i=1}^N
{{\b_i}}\left(V_iV^*_i\rho+\rho V_iV^*_i-2V^*_i\rho V_i\right).
\cr
}
$$
This Lindblad generator admits
$$
\rho_\b={1\over Z_\beta}\,e^{-\beta H_S}
$$
as a unique invariant state and it converges to equilibrium.}

\prf
The announced Lindblad generator is just the dual of the Lindblad
generator described in Corollary 13. 

Let us compute $\rL(\rho_\b)$. We have
$$
[H_S,\rho_\b]=0.
$$
On the other hand, the $V_i$'s have been chosen so that
$$
\rho_\b V_i={{\b_i}\over{\b_0}}V_i\rho_\b.
$$
This gives 
$$
V_i^*\rho_\b={{\b_i}\over{\b_0}}\rho_bV_i^*
$$
These two relation give
$$
V^*_iV_i\rho_\b+\rho_\b V^*_iV_i-2V_i\rho
V^*_i=2V^*_iV_i\rho_\b-2{{\b_0}\over{\b_i}}V_iV_i^*\rho_\b
$$
and
$$
V_iV^*_i\rho_\b+\rho_\b V_iV^*_i-2V^*_i\rho
V_i=2V_iV_i^*\rho_\b-2{{\b_i}\over{\b_0}}V_i^*V_i\rho_\b.
$$
Hence we get the result: $\rL(\rho_\b)=0$.

Let us consider the von Neumann algebra generated by  the
operators $V_i$, $V_i^*$, $i=1\ldots N$. It is easy to see that this is
the whole $\rB(\rH_S)$. Hence the commutant of this von Neumann
algebra is trivial. As we always have the obvious inclusion 
$$
\left\{H_R,V_i,V_i^*;i=1,\ldots
n\right\}'\subset\left\{V_i,V_i^*;i=1,\ldots n\right\}'
$$
we have equality of the two commutants and Theorem 14 applies. This
gives the return to equilibrium property and hence  the uniqueness
of the invariant state.\qed

\bigskip
Note the important following fact: we never used the fact that
$\rho_\b$ is a Gibbs state, we only used the fact that it is a
function of $H_R$. Hence the above result is valid for any state
$\rho$ which is a function of $H_R$.

\spa{}{References}
\noindent[A-W] H. Araki, E.J. Woods: ``Representation of the canonical
commutation relations describing a nonrelativistic infinite free Bose
gas'', {\it Journal of Mathematical Physics} 4 (1963), p. 637-662.
\medskip
\noindent[At1] S. Attal: ``Extensions of the quantum stochastic
calculus'', {\it Quantum Probability Communications} vol. XI, World
Scientific (2003), p. 1-38.
\medskip
\noindent[At2] S. Attal: ``{\it Quantum Noise Theory}'', book to appear,
Springer Verlag.
\medskip
\noindent[AJ1] S. Attal, A. Joye: ``Weak-coupling and continuous limits for
repeated quantum interactions'', {\it preprint}.
\medskip
\noindent[A-L] S. Attal, J.M. Lindsay: ``Quantum stochastic integrals with
maximal domains'', {\it The Annals of
Probability} 32  (2004), p. 488--529.
\medskip
\noindent[AP1] S. Attal, Y. Pautrat: ``From repeated to continuous quantum
interactions'', {\it Annales Henri Poincar\'e (Physique Th\'eorique)},
to appear.
\medskip
\noindent[AP2] S. Attal, Y. Pautrat: ``Nets and continuous products of Hilbert
space. Applications to quantum statistical mechanics'', {\it preprint}.
\medskip
\noindent[Coq] A. Coquio: ``Why are there only 3 quantum noises?'', {\it
Probability Theory and Related Fileds} 118 (2000), p. 349--364.
\medskip
\noindent[Fag] F. Fagnola: `` Quantum Markov semigroups and
quantum flows'', {\it Proyecciones, Journal of Math.} 18 (1999), p. 1-144.
\medskip
\noindent[F-R] F. Fagnola, R. Rebolledo: ``Lectures on the qualitative
analysis of quantum Markov semigroups'', {\it Quantum probability and
white noise analysis}, World Scientific, vol XIV (2002), p.197-240.
\medskip
\noindent[F-V] A. Frigerio, M. Verri: ``Long-time asymptotic
properties of dynamical semigroups on $W^*$-algebras'', {\it
Math. Zeitschrift} (1982).
\medskip
\noindent[G-Z] C.W. Gardiner, P. Zoller: ``{\it Quantum noise. A
handbook of Markovian and non-markovian quantum stochastic methods
with applications to quantum optics}'', 2nd edition, Springer series
in Synergetics, Springer verlag (2000).
\medskip
\noindent[Gui] A. Guichardet: ``{\it Symmetric Hilbert spaces and
related topics}'', Lecture Notes in Mathematics 261, Springer Verlag
(1972).
\medskip
\noindent[H-P] R.L. Hudson, K.R. Parthasarathy: ``Quantum
  It\^o's formula and stochastic evolutions'', {\it Communications in
  Mathematical 
  Physics} 93 (1984), p.  301--323.
\medskip
\noindent[JP1] W. Jaksic, C.-A. Pillet: ``On a model for quantum
friction I: Fermi's golden rule and dynamics at zero temperature'',
{\it Annales Henri Poincar\'e Physique Th\'eo\-ri\-que} 62 (1995),
p. 47-68.
\medskip
\noindent[JP2] W. Jaksic, C.-A. Pillet: ``On a model for quantum
friction II: Fermi's golden rule and dynamics at positive temperature'',
{\it Communications in Mathematical Physics} 176 (1996), p. 619-644.
\medskip
\noindent[JP3] W. Jaksic, C.-A. Pillet: ``On a model for quantum
friction III: Ergodic properties of the spin-boson system'',
{\it Communications in Mathematical Physics} 178 (1996), p. 627-651.
\medskip
\noindent[L-M] J.M. Lindsay, H. Maassen: ``Stochastic calculus
for quantum Brownian motion of nonminimal 
variance---an approach using integral-sum kernel operators'',  
{\it Mark Kac Seminar on Probability and Physics Syllabus 1987--1992}
(Amsterdam, 1987--1992), 97--167,  
 CWI Syllabi, 32, 
 Math. Centrum, Centrum Wisk. Inform., Amsterdam, 1992.
\medskip
\noindent[Par] K.R. Parthasarathy: ``{\it An introduction to quantum
stochastic calculus}'',  Mo\-no\-graphs 
in Mathematics 85, Birkh\"auser (1992).
\end